\documentclass[format=acmsmall, review=false]{acmart}

\usepackage{acm-ec-22}
\usepackage{booktabs} 
\usepackage[ruled]{algorithm2e}

\SetAlFnt{\small}
\SetAlCapFnt{\small}
\SetAlCapNameFnt{\small}
\SetAlCapHSkip{0pt}
\IncMargin{-\parindent}
\setcitestyle{acmnumeric}

\usepackage{amsmath}
\usepackage{amsthm}
\usepackage{url}
\usepackage{mathrsfs}
\usepackage{graphicx}
\usepackage{textcomp}
\usepackage{multirow}
\usepackage{lipsum}
\usepackage{pgfplots}\pgfplotsset{compat=1.16}
\usepackage{subcaption}
\usepackage[export]{adjustbox}
\usepackage{local_macros}

\newtheorem{asm}{Assumption}
\newtheorem{dfn}{Definition}
\newtheorem*{dfn*}{Definition}

\newtheorem{thm}{Theorem}
\newtheorem*{thm*}{Theorem}
\newtheorem{clm}{Claim}
\newtheorem{prop}{Proposition}

\newtheorem*{ex}{Example}

\allowdisplaybreaks

\begin{document}
\title{Order of Commitments in Bayesian Persuasion with Partial-informed Senders}
\author{Shih-Tang Su, Vijay G. Subramanian}

\begin{abstract}
The commitment power of senders distinguishes Bayesian persuasion problems from other games with (strategic) communication. Persuasion games with multiple senders have largely studied simultaneous commitment and signalling settings. However, many real-world instances with multiple senders have sequential signalling. In such contexts, commitments can also be made sequentially, and then the order of commitment by the senders -- the sender signalling last committing first or last -- could significantly impact the equilibrium payoffs and strategies. For a two-sender persuasion game where the senders are partially aware of the state of the world, we find necessary and sufficient conditions to determine when different commitment orders yield different payoff profiles. In particular, for the two-sender setting, we show that different payoff profiles arise if two properties hold: 1) the two senders are willing to collaborate in persuading the receiver in some state(s); and 2) the sender signalling second can carry out a credible threat when committing first such that the other sender's room to design signals gets constrained.
\end{abstract}

\maketitle

\newpage

\section{Introduction} \label{intro}
The cornerstone feature of Bayesian persuasion~\cite{kamenica2011bayesian,rayo:segal:optimalinfo:2010} is the commitment power of the sender(s). Generalized from the seminal prosecutor-judge example in \cite{kamenica2011bayesian} with one sender and one receiver, Gentzkow and Kamanica in \cite{gentzkow2016competition} introduced a more realistic scenario with a second sender, the defense attorney, who can also choose a signal to persuade the judge in addition to the prosecutor; this extended Bayesian persuasion models to multiple senders. A body of literature has since emerged along with two different directions within the context of fully-informed senders: competition and equilibrium under simultaneous signal revelation \cite{au2020competitive}; and (iterative) commitment derivation in sequential persuasion \cite{li2018sequential}. A general result in these works states that the competition (between senders) increases the extent of information revelation.

In a variety of real-world problems with multiple participants but only one decision maker -- such as applying for a grant/scholarship, co-branding marketing, lobbying, and manufacturing process selection -- the senders are not fully informed about the state of the world. However, each sender holds a piece of information about the state of the world, and learning other senders' private information (from their signals) may lead to a more tailored signalling scheme. Hence, truthfully reporting her private information is usually not an optimal strategy for a Bayes-rational sender. The above scenarios can be modeled as a Bayesian persuasion problem with multiple partially informed senders. When senders are partially informed, the interactions of senders under simultaneous revelation and interactions under sequential revelation are essentially different. With sequential signal revelation and partially-informed senders, a sender who sends signals later on in the sequence can learn more about the state of the world by updating her belief using the earlier senders' signals. Since these later on in the sequence  senders can anticipate some revealed signals while sending their own signals, they can design signalling strategies conditional on the realization of early senders' signals. On the other hand, when early senders foresee that their signals will be exploited to act against them, they will modify their commitment to preclude or minimize such behaviors. This iterative belief-updating procedure on commitments leads to the following questions in Bayesian persuasion with multiple partially informed senders context: will the commitment order of senders matter? If it matters, which sender should make her commitment first? In this paper, we study these questions, and start with a motivating example to help the reader better understand real-world scenarios that fit the Bayesian persuasion with multiple partially informed senders paradigm.

\subsection{Motivating Example}
\begin{ex}
A company $X$ searches for the most efficient way to produce its new product. For this purpose, it needs to decide between the following combinations: whether to apply a particular patent or not, and which manufacturing process (advanced or mature) it should adopt. The company sponsors a university lab $Y$ to do some simulations for its decision on the patent application. Based on the simulation results, the company then ask its partner factory $Z$ to do some trial runs to determine the manufacturing process. However, lab $Y$ always prefers the patent to be applied, and the factory $Z$ desires the advanced process to be adopted. Thus, $Y$ may choose simulation environments in favor of the patent being applied, and $Z$ may choose trial run parameters in favor of the advanced process. However, all the simulation and trial-run setups have to be submitted to the company beforehand. Then two questions arise: First, does the order of submission of proposals matter? Second, if it does, from company $X$'s perspective, which party should submit their proposal first, $Y$ or $Z$?
\end{ex}

In the above example, $Y$ and $Z$ play the role of senders with commitments, and $X$ is the receiver. Since both senders only have partial (maybe correlated) information of the state of the world, the sender who sends the signal later, $Z$ in the example, may be able to commit to a signalling scheme that depends on the state of the world and the realized signal of the other sender, $Y$ in the example. Note that when $Z$ makes her commitment, she has not yet observed the signal sent from $Y$. Hence, her goal is to make a commitment on top of $Y$'s commitment (if $Y$ commits first) to exploit the signal revealed from $Y$. However, whether the (potential) exploit of $Y$'s signal can hurt $Y$ depends on the correlation of $Y$ and $Z$'s private information and the relevant payoff matrices. Not only is our example plausible, similar scenarios with multiple partial-informed senders can be found in data markets \cite{acemoglu2019too,fallah2022optimal}, corporate political actions \cite{chiroleu2020merchants}, and stress testing \cite{inostroza2021persuasion}. Hence, 
our goal is to characterize broadly applicable conditions that determine when the commitment order matters. With this goal in mind, the two main contributions of this work are:

\vspace{12pt}
\noindent\textbf{\textit{Contributions:}} 
\begin{enumerate}
    \item To the best of our knowledge, we are the first to identify the importance of the commitment order in multi-sender Bayesian persuasion problems. We show that when senders are partially informed and signals are realized sequentially, the commitment order may play a significant role in the equilibrium strategies (commitments), and hence, the payoffs of the agents. 
    \item In a two-sender setting, we present a set of sufficient conditions and a set of necessary conditions that can identify when the commitment order matters. An intuitive characterization of our conditions is the following -- when both senders have a conflict of interest, and the sender signalling second can post a credible threat, the commitment order will matter.
\end{enumerate}

\subsection{Literature Review}
The major distinction between Bayesian persuasion \cite{kamenica2011bayesian,rayo:segal:optimalinfo:2010} and other paradigms that study games with communication, e.g., cheap talk~\cite{crawford1982strategic}, signalling games~\cite{spence1978job}, and mechanism design \cite{myerson1981optimal,hurwicz2006designing}, is the commitment power of the sender(s). With the commitment power in hand, each sender can influence the receiver's (payoff-relevant) action by committing to her (verifiable) signalling scheme, which is typically a randomized mapping from states of the world to signals. Subsequently, the sender(s) receives some private information about the state of the world and applies the signalling scheme. After that, the receiver(s) choose actions. Seminal works of Kamenica and Gentzkow \cite{kamenica2011bayesian}, Rayo and Segal \cite{rayo:segal:optimalinfo:2010}, and Bergemann and Morris \cite{bergemann2016bayes} established the research area of Bayesian persuasion, and since then there has been a large literature on both theoretical aspects and applications of Bayesian persuasion. To keep our discussion focused, we only discuss closely related work and refer the reader to survey articles~\cite{kamenica2019bayesian,bergemann2019information} for the broader literature.

The literature closest to our work studies Bayesian persuasion with multiple senders. Gentzkow and Kamenica~\cite{gentzkow2016competition} studied the model of fully-informed senders (symmetric information games) with simultaneous signal revelation. They identified conditions on the information environment such that the competition outcome is no less (Blackwell) informative than the collusive outcome, and those conditions were, later on, proved to be sufficient and necessary by Li and Norman~\cite{li2018bayesian}. Thereafter, Gentzkow and Kamenica~\cite{gentzkow2017bayesian} and Au and Kawai~\cite{au2020competitive} both considered fully-informed senders with simultaneous signal revelation. The former proved that the concavification approach could be extended to multiple senders, and greater competition among senders tends to increase information revelation for the receiver. The latter further showed the existence of a unique symmetric equilibrium, and also that full disclosure (of every sender) occurs when the number of senders goes to infinity. 
In simultaneous commitment plus signal revelation models, the competition among senders becomes more sophisticated when senders are partially informed. In \cite{boleslavsky2018limited}, Boleslavsky and Cotton studied two senders with independent partial information. They construct the equilibrium based on the result that each sender’s incentive is similar to that in an all-pay auction under complete information. Au and Kawai~\cite{au2021competitive} considered multiple senders with correlated information, and presented two effects, namely the underdog-handicap effect and the good-news curse, that drive the information disclosure. The former encourages more aggressive disclosure because of the receiver's bias towards a (ex-ante) stronger sender. But the latter disincentivizes the disclosure because each sender using her favorable
signal implies that their rival is more likely to generate a strong competing signal in response. However, when senders are partially informed, interactions among senders can incorporate not only competitions but also (partial) collaboration. The simultaneous commitment plus signal revelation models essentially preclude\footnote{Except for senders with fully-aligned utilities, which will transform the multi-sender Bayesian persuasion problem to a coordination game.} collaboration among partially informed senders since senders cannot learn (and tailor) their commitments using other senders' possible realized signals. Hence, to fully capture the interactions among partially-informed senders, one should target settings beyond the simultaneous signal revelation paradigm. 

Besides the simultaneous signal revelation setting, Li and Norman~\cite{li2018sequential} consider a Bayesian persuasion model with sequential signal revelation. In the Li-Norman model, senders commit sequentially to post experiments, and every experiment is conditionally independent of each other given the state. With this conditional independence, Li and Norman analyze optimal signalling strategies for heterogeneous\footnote{Senders with different utility functions.} senders with same (and full\footnote{The sender can access all possible experiments under the ``reduced'' state space obtained by merging each state set not distinguishable by the sender to a single state.}) information about the state. Interestingly, their result shows that sequential persuasion cannot yield a more informative equilibrium in comparison to simultaneous persuasion \cite{gentzkow2016competition} and is strictly less informative in binary state models. Their result stands on the conditional independence of the experiments and the assumption that the order of (senders') commitments perfectly matches the order of signals/experiments. However, in scenarios such as the motivating example above, this alignment may not hold, and the receiver may have a choice about who should commit first. Besides, analogous to simultaneous Bayesian persuasion, senders in sequential Bayesian persuasion may be partially informed. The sequential revelation and partially informed senders assumptions distinguish our model from the existing literature, and leads to interesting follow-up research questions on the optimal commitment order.
\section{Problem Formulation}
There are two senders $\Sender{1},\Sender{2}$ and one receiver $R$ in the game. The state space $\StateSpace$ is finite, and the receiver's action space $\ActSpace$ is also finite. For simplicity of analysis, we assume that the state and action space have the same size, i.e., $|\StateSpace|=|\ActSpace|$, and action $\Act{\STATE}$ is the receiver's unique best response of state $\STATE\in\StateSpace$. To further simplify the analysis, we assume that the receiver only obtains (positive) utility upon matching the state correctly, i.e., 
\begin{eqnarray}
 \Util{R}{\StateX{i}}{\Act{j}}
 \begin{cases}
 1 & \text{if~} i=j,\\
 0 & \text{otherwise}.
 \end{cases}
\end{eqnarray}
Before the true state is realized (by nature), the senders and the receiver have a common prior belief of the distribution of the state $\PRIOR\in \Delta(\StateSpace)$. However, both senders obtain some private information about the state of the world when it is realized. To avoid some trivial cases and also to avoid overlap with the existing literature, we make four assumptions on the senders' private information:
\begin{asm}
Each sender only has partial information of the true state. To avoid redundancy, we assume each sender \Sender{k}'s information space $\Infoset{k}$ has the size $|\Infoset{k}|<|\StateSpace|$, $k\in\{1,2\}$.
\end{asm}
\begin{asm}
    Senders $\Sender{1}$ and $\Sender{2}$ have different private information, i.e., $\Infoset{1}\neq \Infoset{2}$.
\end{asm}

\begin{asm}
Neither sender is more knowledgeable\footnote{In terms of Blackwell informativeness~\cite{blackwell1953equivalent}.} than the other, and the private information of each sender is generated via an onto function $F_k: \StateSpace \rightarrow \Infoset{k}$. In other words, $F_k$ is a partition of $\StateSpace$, and neither $F_1$ is more Blackwell informative than $F_2$ nor $F_2$ is more Blackwell informative than $F_1$.
\end{asm}
\begin{asm}
The true state can be revealed under both senders' truth-telling strategies. In other words, for every $\StateX{i} \in \StateSpace$, there exists an $\Info{1}{x}\in\mathcal{I}_1$ and an $\Info{2}{y}\in\mathcal{I}_2$ such that $\Prob{\StateX{i}|\Info{1}{x},\Info{2}{y}}=1$.    
\end{asm}

After both senders get their private information, \Sender{1} sends a signal $\Signal{1} \in\SignalSpace_1$ to influence the receiver's belief. After observing the signal \Signal{1}, \Sender{2} sends a signal $\Signal{2} \in\SignalSpace_2$ to influence the receiver's belief. After receiving signals \Signal{1} and \Signal{2}, the receiver has to take action $\ACT\in \ActSpace$. Extending the sender-preferred tie-breaking rules in \cite{kamenica2011bayesian}, we assume that the receiver breaks a tie in favor of \Sender{1} first. When the action set maximizing both \Sender{1} and receiver's utilities is not a singleton, the receiver takes an (arbitrary) action maximizing \Sender{2}'s utility from the set.

In this model, both senders commit to their respective signalling strategies before receiving their private information. Moreover, \Sender{1} and \Sender{2} make their commitment sequentially under a pre-determined order. That is to say, the sender who commits later can exploit another sender's commitment to design her signalling strategy. We will first detail the game flow (when the order of commitment in steps 2 and 3 is determined), and then discuss how the sender's commitment can depend on the other sender's signalling strategy based on the commitment order.

\subsection{Procedure of the game} 
The game in our model evolves as follows:
\begin{enumerate}
    \item Prior belief of the state $\PRIOR\in \Delta(\StateSpace)$ is disclosed to all participants.
    \item \Sender{k} commits her signalling strategy, where the index $k\in\{1,2\}$ is determined beforehand and becomes common knowledge.
    \item The other sender, \Sender{-k}, commits to her signalling strategy.
    \item Nature chooses a realized state $\STATE \in \StateSpace$.
    \item \Sender{1} gets her private information and sends a (randomized) signal $\Signal{1}$ based on her commitment.
    \item \Sender{2} gets her private information and observes $\Sender{1}$'s realized signal \Signal{1}. Then, she sends a (randomized) signal $\Signal{2}$ according to her commitment.
    \item The receiver observes the realized signals $\Signal{1},\Signal{2}$ and takes action $\ACT\in A$ to maximize the probability of matching the state.
\end{enumerate}    

\subsection{Receiver's best response}
    Given the procedure of the game, the receiver takes action after observing the realization of both senders' signals, $\Signal{1},\Signal{2}$. Given that her objective is to match the state correctly, the receiver's best response is taking action $\ACT^*\in \ActSpace$ such that
    \begin{align*}
        \ACT^*\in \arg\max_{\ACT\in \ActSpace}\Expect{p,\Gamma_1,\Gamma_2}{\Util{R}{\STATE}{\ACT}|\Signal{1},\Signal{2}}.
    \end{align*}
    However, the tie-breaking rule, although indifferent to the receiver, plays a significant role on senders' signalling strategy, especially when multiple senders are considered in the model. As mentioned above, we assume the receiver breaks a tie in favor of $\Sender{1}$ first and then \Sender{2}. The formal expression of the tie-breaking rule is presented below: 
    \begin{align}
        &\text{Let}~\ActSpace_{R}\coloneqq\arg\max_{\ACT\in \ActSpace}\Expect{p,\Gamma_1,\Gamma_2}{\Util{R}{\STATE}{\ACT}|\Signal{1},\Signal{2}} \text{,~} \ActSpace_{\Sender{1}}\coloneqq\arg\max_{\ACT'\in \ActSpace_{R}} \Expect{p,\Gamma_1,\Gamma_2}{\Util{\Sender{1}}{\STATE}{\ACT'}|\Signal{1},\Signal{2}},\nonumber\\
        &\Expect{p,\Gamma_1,\Gamma_2}{\Util{\Sender{1}}{\STATE}{\ACT^*}|\Signal{1},\Signal{2}}=\max_{\ACT'\in \ActSpace_{R}} \Expect{p,\Gamma_1,\Gamma_2}{\Util{\Sender{1}}{\STATE}{\ACT'}|\Signal{1},\Signal{2}},\label{eqn:4.3}\\
        &\Expect{p,\Gamma_1,\Gamma_2}{\Util{\Sender{2}}{\STATE}{\ACT^*}|\Signal{1},\Signal{2}}=\max_{\ACT''\in \ActSpace_{\Sender{1}}} \Expect{p,\Gamma_1,\Gamma_2}{\Util{\Sender{2}}{\STATE}{\ACT'}|\Signal{1},\Signal{2}}. \label{eqn:4.4}
    \end{align}
To avoid ambiguity, $\Expect{p,\Gamma_1,\Gamma_2}{\UTIL{}(\STATE,a)|\Signal{1},\Signal{2}}$ represents the expected utility conditional on the realized signals $\Signal{1}$ and $\Signal{2}$ under the prior $p$, \Sender{1}'s commitment $\Gamma_1$, and \Sender{2}'s commitment $\Gamma_2$, given action $a$ is taken. In short, the expectation is taken only on the prior. Equation (\ref{eqn:4.3}) states that the receiver chooses an (arbitrary) action in the set which maximizes \Sender{1}'s expected utility while she is indifferent. Equation (\ref{eqn:4.4}) states if there is still a tie after maximizing both \Sender{1} and receiver's expected utility, the receiver chooses an (arbitrary) action in the set which maximizes \Sender{2}'s expected utility. For simplicity of representation, we abuse notation and let $\ACT^*(\Signal{1},\Signal{2})$ represents the receiver's best response under $\Signal{1}$ and \Sender{2}'s signal $\Signal{2}$ hereafter.

\subsection{Assumption on commitments} \label{sec:commitassume}
After defining the receiver's best response, we detail why a sender \Sender{k} may be able to utilize another sender \Sender{-k}'s commitment when she commits later. The key is the pre-determined order of the sequential signalling. Since \Sender{1} always sends her signal before \Sender{2}, the signalling schemes of the senders are asymmetric. In other words, \Sender{1} cannot commit to a signalling scheme that depends on the realization of \Signal{2}, but \Sender{2} can commit to a signalling scheme based on the realization of \Signal{1}. Next we clarify the information that \Sender{2} can exploit while making her commitment. To avoid ambiguity, we discuss both commitment orders and will introduce our cornerstone assumption, namely \textbf{permutation-free} commitments. Before the discussion, we note that the superscript $f,s$ of the commitment notation $\Gamma^{f},\Gamma^{s}$ denote the commitment order, and the subscript denote the sender and another sender's commitment (if available), e.g., $\Gamma^s_{2,\Gamma^f_1}$ denotes \Sender{2}'s commitment when she commits after \Sender{1} under \Sender{1}'s commitment $\Gamma^f_1$.
\vspace{3pt}\\
\textbf{Case 1: \Sender{1} commits to her signalling strategy first}
\begin{itemize}
    \item Since sender \Sender{1} has no observation of sender \Sender{2}'s signal while sending her signal and has to commit first, her commitment $\Gamma^f_1$ can be represented as a function $\Gamma^f_1: \Infoset{1}\rightarrow \Delta(\SignalSpace_1)$, where $|\SignalSpace_1|\leq |\Infoset{1}|$.
    \item Sender $S_2$ can utilize sender $S_1$'s commitment and realized signal to make her commitment. Hence, sender $S_2$'s commitment $\Gamma^s_{2,\Gamma^f_1}$ can be represented as a function $\Gamma^s_{2,\Gamma^f_1}:\Infoset{2}\times \SignalSpace_1 \rightarrow \Delta(\SignalSpace_2)$.
\end{itemize}
\vspace{3pt}
\textbf{Case 2: $S_2$ commits to her signalling strategy first}
\vspace{-6pt}
\begin{itemize}
    \item Even though sender \Sender{1} knows \Sender{2}'s commitment $\Gamma^f_2$, she still has no observation of sender \Sender{2}'s signal while sending her signal. Hence, the commitment $\Gamma^s_{1,\Gamma^f_2}$ is a function $\Gamma^s_{1,\Gamma^f_2}:\mathcal{I}_1\rightarrow \Delta(\SignalSpace_1)$, where $|\SignalSpace_1|\leq |\mathcal{I}_1|$. In other words, sender \Sender{1} can adjust her commitment based on \Sender{2}'s commitment, but cannot tailor her commitments by utilizing \Sender{2}'s realized signals.
    \item Sender \Sender{2} wants to utilize \Sender{1}'s realized signal to make her (optimal) commitment. However, \Sender{2} has to commit before sender \Sender{1} in this case. Since we don't want each signal token to possess an implied meaning, thereby allowing \Sender{1} to act against \Sender{2} by just reordering the signal tokens, we assume that the \Sender{2} commits to a \textbf{permutation-free commitment}. 
\end{itemize}

\begin{dfn}
A commitment $\Gamma^f_{2}:\Infoset{2}\times \SignalSpace_1 \rightarrow \Delta(\SignalSpace_2)$ is \textbf{permutation-free} if for every commitment $\Gamma^s_{1}$ of \Sender{1}, there is no permutation matrix $M$ such that
\begin{align*}
    \Expect{p,\Gamma^s_{1}}{\UTIL{\Sender{2}}|\Gamma^f_{2}(I_2,M(\Signal{1}))}> \Expect{p,\Gamma^s_{1}}{\UTIL{\Sender{2}}|\Gamma^f_{2}(I_2,\Signal{1})}.
\end{align*}
\end{dfn}

Next, we justify why we believe that the permutation-free assumption on commitments is appropriate when \Sender{2} has to commit first. When we exploit Bayesian persuasion in real-world problems, commitments are usually implemented via experiments (which can be operated/supervised by a third party). Hence, when \Sender{2} has to commit first and wants to exploit \Sender{1}'s realized signal, she can propose different experiments depending on \Sender{1}'s signal realization. Unless we assume some signal tokens have implied meaning or \Sender{2} has some prior knowledge about \Sender{1}'s signalling strategy, \Sender{2} cannot know the mapping from signal \Signal{1} to the interim distribution of the states while making her commitments. However, when an experiment of \Sender{2} has to be executed, \Sender{1}'s signal realization and her commitments are both common knowledge. Hence, \Sender{2} can commit to experiments with her private signal and the mock signal realizations of \Sender{1}, and ask the operator of her experiments to use a permutation/reordering of \Sender{1}'s signal tokens that works best for her. These experiments can be appropriately executed, and this permutation-free commitment can be fulfilled. Moreover, not embracing the permutation-free property implies that \Sender{1} and \Sender{2} have an \textit{ex-ante} consensus on the meaning of \Sender{1}'s signal tokens, but \Sender{1} can violate their ex-ante consensus to act against \Sender{2}. If \Sender{2} can lose by sender \Sender{1} permuting her signal tokens, \Sender{2}'s experiments will reveal her (partial) private information only if \Sender{1} cannot exploit this revelation against her via permutations. This will significantly reduce the utility of being able to choose the commitment order. To further illustrate this, a numerical example is presented in Appendix \ref{app:permutation_free} where \Sender{2}'s optimal commitment ends up being independent of \Sender{1}'s signal realization when the permutation-free property is violated. 

\subsection{Objectives of senders} \label{sec:objIV}

Before we proceed, we list the objective functions of senders for both commitment orders. For simplicity of representation, we abuse notations and let $\Gamma_k^{f*}$ denotes the optimal commitment when \Sender{k} commits first, and $\Gamma_k^{s*}$ denotes the optimal commitment when \Sender{k} commits last.

When \Sender{1} commits first, the objective functions of \Sender{1} and \Sender{2} are the following:
\begin{align}
    \Gamma_1^{f*}&\in\arg\max_{\Gamma_1^f\in\mathbf{\Gamma_1}}\Expect{p,\Gamma_1^{f},\Gamma_2^s(\Gamma_1^f)}{U_{S_1}} \nonumber\\ &\text{~s.t.~}  \Expect{p,\Gamma_1^{f},\Gamma_2^s(\Gamma_1^f)}{U_{S_2}}=\max_{\Gamma_2^{s'}\in\mathbf{\Gamma_2}} \Expect{p,\Gamma_1^{f},\Gamma_2^{s'}}{U_{S_2}}, \\
    \Gamma_2^{s*}&\in\arg\max_{\Gamma_2^s\in\mathbf{\Gamma_2}}\Expect{p,\Gamma_1^{f*},\Gamma_2^s}{U_{S_2}}.
\end{align}

When \Sender{2} commits first with permutation-free commitments, the objective functions of \Sender{1} and \Sender{2} are the following:
\begin{align}
    \Gamma_1^{s*}&\in\arg\max_{\Gamma_1^s\in\mathbf{\Gamma_1}}\Expect{p,\Gamma_1^s,\Gamma_2^{f*}}{U_{S_1}}\text{~s.t.~}  \Expect{p,\Gamma_1^{s},\Gamma_2^{f*}}{U_{S_2}}=\max_{\Gamma_1^{s'}\in \mathbf{\Gamma_1^M}(\Gamma_1^{s})}\Expect{p,\Gamma_1^{s'},\Gamma_2^{f*}}{U_{S_2}},  \nonumber\\ &\text{~where~} \mathbf{\Gamma_1^M}(\Gamma_1^{s}) \text{~is the set of commitments where the signals are permuted} \nonumber\\ &~\text{when compared to~} \Gamma_1^{s}. \\
    \Gamma_2^{f*}&\in\arg\max_{\Gamma_2^f\in\mathbf{\Gamma_2}}\Expect{p,\Gamma_1^s,\Gamma_2^{f}}{U_{S_2}} \text{~s.t.~}  \Expect{p,\Gamma_1^{s},\Gamma_2^f}{U_{S_1}}=\max_{\Gamma_1^{s'}\in\mathbf{\Gamma_1}} \Expect{p,\Gamma_1^{s'},\Gamma_2^{f}}{U_{S_1}}. 
\end{align}

\section{When the Commitment Order Matters -- Simple conditions}
In Section \ref{sec:commitassume}, we discussed the rationale for why the permutation-free property (on commitments) is an appropriate assumption in sequential commitment problems. In this section, we will present an intuitive set of conditions when the commitment order matters, i.e., when different expected utilities are obtained for different commitment orders so that senders or the receiver would have a preference between them. We start by presenting a numerical example that shows that the commitment order can result in a credible threat even though the order of signalling stays unchanged.

\subsection{Credible threats} \label{sec:4.3.1}

Let's consider a game with two senders and one receiver. There are 4 possible states of the world $\SignalSpace=\{TL,TR,BL,BR\}$ (top-bottom and left-right). \Sender{1} knows whether the state of the world is top or bottom, i.e., $\Infoset{1}=\{T,B\}$, and \Sender{2} knows whether the state of the world is left or right, i.e., $\Infoset{2}=\{L,R\}$. \Sender{1} can send signals $\Signal{1}\in \{\Signal{1,1},\Signal{1,2}\}$ and \Sender{2} can send signals $\Signal{2}\in \{\Signal{2,1},\Signal{2,2}\}$ to influence the receiver's action. The distribution of the state of the world (prior) is common knowledge, listed in Table~\ref{tb5}. The senders' utilities are provided in Table~\ref{tb6}. The receiver's goal is to guess the state of the world accurately: she receives payoff 1 when she guesses correctly and 0 otherwise.

\begin{table}[ht]
 \caption{State distribution and senders' utilities}
   \label{tb:example_credible_threat}
    \vspace{-8pt}
  \begin{subtable}[t]{0.48\textwidth}
  \centering
\begin{tabular}{|l|l|l|} 
\hline
  & L   & R   \\ \hline
T & $0.1$ & $0.2$ \\ \hline
B & $0.4$ & $0.3$ \\ \hline
\end{tabular} 
\caption{Distribution of the states} \label{tb5}
  \end{subtable}
  \begin{subtable}[t]{0.48\textwidth}
  \centering
\begin{tabular}{|l|l|l|l|l|} 
\hline
  & $\Act{TL}$   &   $\Act{TR}$ &   $\Act{BL}$ &   $\Act{BR}$\\ \hline
$(U_{S_1},U_{S_2})$ & 1,0 & 2,2 & 0,0 & 0,3 \\ \hline
\end{tabular}  
\caption{Utilities of senders} \label{tb6}
\end{subtable}
\vspace{-1pt}
\end{table}


In this game, \Sender{1} prefers top over bottom and \Sender{2} prefers right over left. Now, we study the optimal commitments of \Sender{1} and \Sender{2} under different commitment orders. First, let's start with the scenario where \Sender{1} make her commitment first.

\subsubsection{\texorpdfstring{\Sender{1}}{Sender 1} commits first}

    If \Sender{2} didn't appear in the game, \Sender{1} would act as if in a classical 1-sender Bayesian persuasion model: sending a signal which suggests action $\Act{TL}$ or $\Act{TR}$ when she knows the true state is in the top half and perform a mixed signalling strategy when the true state is in the bottom half. However, in the presence of \Sender{2}, \Sender{1} knows that as long as she can guarantee that the posterior probability of $BR$ is smaller than or equal to $TR$, \Sender{2}, for the sake of maximizing her expected payoff, will commit to a signalling strategy which helps \Sender{1} since $BL$ is inferior to $TR$ for \Sender{2}. Bearing this in mind, \Sender{1} will make the following commitment by anticipating \Sender{2}'s commitment.
    \vspace{3pt}\\
    \textbf{\Sender{1}'s commitment}: Send a suggestion of $T$ with probability 1 when her private information is $T$ and probability $\frac{2}{3}$ when her private information is $B$.

    Now, since \Sender{1} has made her commitment, \Sender{2} makes her commitments on top of it.  \vspace{3pt}\\
\textbf{\Sender{2}'s commitment:}   
    \begin{itemize}
\item While observing \Sender{1}'s signal suggesting $T$, send a suggestion $R$ with probability 1 when her private information is $R$ and probability $\frac{3}{4}-\epsilon$ when her private information is $L$. And send a suggestion of $L$ with probability $\frac{1}{4}+\epsilon$ while observing $L$, where $\epsilon$ is infinitesimal, i.e., $|\epsilon| \ll 1$.
\item  While observing \Sender{1}'s signal suggesting $B$, send a suggestion $R$ with probability 1 when her private signal is $R$ and probability $\frac{3}{4}-\epsilon$ when her private signal $L$. And send a suggestion of $L$ with probability $\frac{1}{4}+\epsilon$ while observing $L$, where $\epsilon$ is infinitesimal, i.e., $|\epsilon| \ll 1$.
\end{itemize}

We notice that \Sender{2}'s commitment is independent of \Sender{1}'s realized signal. This is merely a happenstance in this example; this makes the analysis simpler too. Given the commitments above, the expected utilities of \Sender{1},\Sender{2} and the receiver are $[\Expect{\PRIOR}{\UTIL{\Sender{1}}}, \Expect{\PRIOR}{\UTIL{\Sender{2}}},\Expect{\PRIOR}{\UTIL{R}}] =[1.2,1.8,\tfrac{1}{3}]$\footnote{See Appendix \ref{app:utilcal3.1.1} for detailed calculation.}. 

\subsubsection{\texorpdfstring{\Sender{2}}{Sender 2} commits first}
Now, consider the game where \Sender{2} commits first. Although \Sender{2} does not know \Sender{1}'s signalling strategy while making her commitment, her commitment can still use \Sender{1}'s signal realization. This is because, when her experiment is conducted, \Sender{1}'s signal realization is common knowledge. Therefore, she can commit to the following signal and use this to threaten \Sender{1} credibly. \vspace{15pt}\\
\textbf{\Sender{2}'s commitment}
\begin{itemize}
\item While observing \Sender{1}'s suggestion of $T$, send a suggestion $R$ with probability 1, whatever her private information is.
\item While observing \Sender{1}'s suggestion of $B$, send a suggestion $R$ with probability 1 when her private information is $R$ and probability $\frac{3}{4}$ when her private information is $L$. And send a suggestion $L$ otherwise.
\end{itemize}

Given this commitment of \Sender{2}, \Sender{1}, now, cannot hope that \Sender{2} will help her when \Sender{1} signals her private information is $T$. Hence, \Sender{1} has to make a more conservative commitment. \vspace{3pt}\\
\textbf{\Sender{1}'s commitment}
\begin{itemize}
    \item Send a suggestion $T$ with probability 1 when her private information is $T$ and probability $\frac{1}{2}$ when \Sender{1}'s private information is $B$.
    \item Send a suggestion $B$ otherwise.
\end{itemize}
    
    Now, we have to check whether \Sender{2} will deviate from her original commitment to verify optimality from \Sender{2}'s perspective. Since the receiver breaks a tie in favor of \Sender{1} and then \Sender{2}, using any signal $L$ with probability $>0$ when $\Sender{1}$ sends a suggestion $T$ will strictly lower her expected utility. When $\Sender{1}$ suggests $B$, the true state is either $BL$ or $BR$, and her commitment already maximizes her expected utility by making the receiver weakly prefer $\Act{BR}$ (given the receiver's tie-breaking rule and the indifference of \Sender{1}'s utility between $\Act{BL}$ and $\Act{BR}$). Hence, her original commitment is the best response to the \Sender{1}'s commitment.
    By calculating\footnote{See Appendix \ref{app:utilcal3.1.2} for detailed calculation.} the expected utility of agents, we obtain: $[\Expect{\PRIOR}{\UTIL{\Sender{1}}}, \Expect{\PRIOR}{\UTIL{\Sender{2}}},\Expect{\PRIOR}{\UTIL{R}}] =[1.1,3,0.4]$.

\subsubsection{Comparison of utilities}    
    Next, we compare the utilities for senders and the receiver between the two commitment orders and assess their value for each agent. Both senders want to commit first, and the receiver has a preference on letting \Sender{2} commit first. Moreover, if senders can trade their commitment order and \Sender{1} is initially endowed the first commitment slot, every transfer $x\in(0.1,0.2)$ from \Sender{2} to \Sender{1} makes both senders better off by trading the first commitment slot from \Sender{1} to \Sender{2}. The example shows an interesting phenomenon: when senders only have partial information and have to send signals sequentially, they may collaborate on the signals they commit and collude on the order of commitment if there is a market for senders to trade their commitment position. For readers curious about whether there is a scenario that both senders and the receiver are better off after a transfer of the commitment slot, an example is provided in Section \ref{sec:magaex}.

\subsection{Collaborative States} \label{sec:commitresult}
    Motivated by the above example, we first present an easily rationalized proposition capturing a set of scenarios where the commitment order matters and prove a generalized theorem afterward. The proposition captures scenarios similar to the example discussed above, wherein \Sender{1}'s and \Sender{2}'s interests are aligned if \Sender{1} reveals her private information truthfully. In these scenarios, both senders want to persuade the receiver to take the same action. This implies that senders can collaborate on their signalling strategies in some states. We call this type of states \textbf{collaborative states} and define them formally below.
\begin{dfn} \label{def:collaborative}
A state $\hat{\STATE} \in \StateSpace$ is a collaborative state if the following two conditions hold:
\begin{enumerate}
    \item $\hat{\STATE} \in \arg\max_{\STATE\in \hat{I}_1}\Util{S_1}{\STATE}{\Act{\STATE}}$,
    \item $\hat{\STATE} \in \arg\max_{\STATE\in \hat{I}_1}\Util{S_2}{\STATE}{\Act{\STATE}}$,
\end{enumerate}
 where $\hat{I}_1$ is the information set containing state $\hat{\STATE}$, i.e., $\hat{I}_1=F_1(\hat{\STATE})$.
\end{dfn}

Definition \ref{def:collaborative} states that when the \Sender{1} reveals her private information truthfully while $\STATE=\hat{\STATE}$, \Sender{2} is willing to collaborate with \Sender{1} to persuade the receiver towards taking their most preferred action. Note the asymmetry of the definition vis-a-vis the sender identities. This occurs because \Sender{1} signals before \Sender{2}. With the definition of collaborative states in place, we present a preliminary result about conditions when the commitment order matters.

\begin{prop} \label{lem:prelim}
Given a prior $p$ and both senders' utility functions $\UTIL{S_1},\UTIL{S_2}$, the commitment order matters if there exists a private information set $I_1\in\Infoset{1}$ satisfying the following conditions:
\begin{enumerate}
    \item (\textbf{Existence of collaborative state}) There exists a collaborative state $\hat{\STATE}\notin I_1$ satisfying the following two inequalities:
    \begin{align}
        &\Util{\Sender{2}}{\hat{\STATE}}{\Act{\hat{\STATE}}}<
        \Expect{p,\Gamma_2^*(I_1)}{\Util{\Sender{2}}{\STATE}{\ACT^*(I_1,\Signal{2})}|I_1} \label{eq:4.8}  \\
     &\Util{\Sender{1}}{\hat{\STATE}}{\Act{\hat{\STATE}}}> \Expect{p,\Gamma_2^*(I_1)}{\Util{\Sender{1}}{\STATE}{\ACT^*(I_1,\Signal{2})}|I_1},  \label{eq:4.9}  
    \end{align}
     where $\Gamma_2^*(I_1)$ denotes \Sender{2}'s optimal signalling strategy while \Sender{1} truthfully reveals her private signal $I_1$, and $\ACT^*(I_1,\Signal{2})$ denotes the receiver's best response under $I_1$ and \Sender{2}'s signal $\Signal{2}$.
    \item (\textbf{A demand of collaboration}) Let $\mathbf{\hat{\Gamma}}_{1}$ be the set of commitments with a signal $\hat{\omega}_1\in\SignalSpace_1$ satisfying $\mathbb{P}(I_1|\hat{\omega}_1)=1$. For any commitment order, the optimal commitment(s) of \Sender{1} must belong to this set, i.e., $\Gamma_1^{f*},\Gamma_1^{s*} \subseteq \mathbf{\hat{\Gamma}}_{1}$.
    \item (\textbf{Existence of a credible threat}) The preference orders of \Sender{1} and \Sender{2} on $\Phi(I_1)$ are polar opposites, where $\Phi(I_1)$ is the receiver's possible optimal action set given \Sender{1}'s private information $I_1$, i.e., $\Phi(I_1)=\{\Act{\STATE}|\STATE\in I_1\}$.
\end{enumerate}
\end{prop}
We provide a detailed proof in Appendix \ref{pf:prop_prelim}. The core of the proof is to construct a credible threat using a collaborative state. To make the construction possible, we need the existence of another signal which gives $\Sender{1}$ a lower expected utility (based on the receiver's best response). This other signal gives \Sender{1} an incentive to commit to a mixed signalling strategy, which mixes this signal with the signal corresponding to the collaborative state. Otherwise, \Sender{1}'s private information can be elicited by \Sender{2} when \Sender{1}'s private information contains the collaborative state, and credible threats are no longer needed since \Sender{1}'s and \Sender{2}'s interests are aligned. The existence of this other signal is implied by condition 2 in non-trivial scenarios, where the details are stated in Claim \ref{clm:4}. Moreover, conditions 1 and 3 guarantee a conflict of interest between \Sender{1} and \Sender{2}'s on mixing other states with the collaborative state. Hence, if \Sender{1} gets to commit first, she can propose a more aggressive mixed strategy because the best commitment of \Sender{2}, after \Sender{1}'s commitment, is to collaborate with \Sender{1}'s mixed strategy. However, if \Sender{2} can commit first, she is willing to collaborate, but she does not want \Sender{1} to mix other states with this collaborative state. Hence, she can create a credible threat to reduce (but not eliminate) \Sender{1}'s strategy of mixing other states with the collaborative one. Hence, the commitment order matters.

\begin{clm} \label{clm:4}
When the inequality (\ref{eq:4.9}) holds, condition 2 in Proposition \ref{lem:prelim} implies that one of the following two conditions is true:
\begin{enumerate}
    \item The signal $\hat{\omega}_1$ of condition 2 in Proposition 4 is not the signal realization which gives \Sender{1} the highest expected utility in \Sender{1}'s optimal commitments.
    \item Every signal realization gives \Sender{1} the same expected utility.
\end{enumerate}
\end{clm}

\section{General Results On When The Commitment Order Matters} \label{sec:general}

Note that while Proposition \ref{lem:prelim} provides a set of sufficient conditions when the commitment order matters in a constructive manner, each condition in Proposition \ref{lem:prelim} has its deficiency. First, the requirement of a collaborative state is a strong condition. \Sender{2} may be willing to partially collaborate with \Sender{1} after she has maximized the probability of her most-preferred action under a given $I_1$. In other words, a collaboration may happen in a state where its corresponding action is \Sender{2}'s second-best action (under the given information set of \Sender{1}). Second, the verification of demand of collaboration requires knowledge about the commitment space's structure\footnote{When the utilities can be written in a matrix form, it is usually straightforward to see whether there is an information set of \Sender{1} that gives her a lower utility compared to the collaborative state.}. When the state space grows, this can be a tedious task without additional constraints. Third, although the credible-threat condition guarantees the uniqueness of \Sender{2}'s optimal signalling strategy and a conflict of interest between \Sender{1} and \Sender{2} on $I_1$, a polar opposite preference ordering is only one type where a conflict of interest between \Sender{1} and \Sender{2} happens. As long as \Sender{1} and \Sender{2} have an opposite preference on a mixture of two information sets of $\Sender{1}$, \Sender{2} may be able to construct a credible threat using this opposite preference. Before stating the generalized result, we start with an example where the commitment order matters but the conditions in Proposition \ref{lem:prelim} are violated in Section \ref{sec:magaex}. Hence, to remove the need for the knowledge of the commitment space, the requirement of collaborative states, and the assumption of polar opposite preference ordering, we will present a set of sufficient conditions generalized from Proposition \ref{lem:prelim}. 

\subsection{An example highlighting the lack of generality of Proposition~\ref{lem:prelim}} \label{sec:magaex}
As mentioned above, we will present an example where the commitment order matters, but conditions in Proposition \ref{lem:prelim} are either violated or not needed. Besides, we utilize this example to demonstrates a scenario where transferring the commitment slot increases the utilities of both senders and also the receiver, i.e., the social welfare increases. In this example we highlight that a collaborative state is not necessary for \Sender{1} to mix her private information sets. Besides, a conflict of interest between \Sender{1} and \Sender{2} on a mixture of a pair of \Sender{1}'s information sets does not demand (polar) opposite preference ordering (on corresponding actions) in any of \Sender{1}'s private information sets. 

Although the smallest example that meets our need requires 7 states, to simplify the description of $\Sender{1}$'s and $\Sender{2}$'s private information structures, we add two dummy states with prior probability $0$ and expand the example to a $9$-state model, where each sender has three possible types. To avoid confusion and to simplify the discussion on sender's preference ordering, we skip the definition of \Sender{1}'s and \Sender{2}'s utilities on dummy states.

\begin{ex} \label{ex:complex}
There are $3\times 3=9$ states of the world $\StateX{XY}\in \StateSpace$, where $X\in \{T,M,B\}$ and $Y\in \{L,C,R\}$. \Sender{1} knows the information of $X$ and \Sender{2} knows the information of $Y$, i.e., $\Infoset{1}=\{T,M,B\}$ and $\Infoset{2}=\{L,C,R\}$. The prior on the state is provided in Table \ref{tb7}, and the utilities of \Sender{1} and \Sender{2} are provided in Table \ref{tb8}. In this example, \Sender{1} sends signals prior to \Sender{2}, and the receiver's objective is to maximize the probability of guessing the true state. To make notation less cumbersome, we use $\Act{XY}$ to represent $\Act{\StateX{XY}}$.

\begin{table}[ht] 
	\centering
 \caption{State distribution and senders' utilities}
 \vspace{-8pt}
\begin{subtable}[t]{0.38\textwidth}
	\centering
\begin{tabular}{|l|l|l|l|}  
\hline
  & L  & C & R   \\ \hline
T & $0.05$ & $0$ & $0.05$ \\ \hline
M & $0.12$ & $0.2$ & $0.02$ \\ \hline
B & $0.35$ & $0.21$ & $0$ \\ \hline
\end{tabular}
	\caption{Distribution of the states}  \label{tb7}
\end{subtable}
\begin{subtable}[t]{0.58\textwidth}
	\centering
\begin{tabular}{|l|l|l|l|l|l|l|l|l|} 
\hline
  & $\Act{TL}$  &   $\Act{TR}$ 
  & $\Act{ML}$  &   $\Act{MC}$  &   $\Act{MR}$
  & $\Act{BL}$  &   $\Act{BC}$\\ \hline
$(U_{S_1},U_{S_2})$ & 0,0 & 5,5 & 1,0 & 2,2 & 0,3
& 0,1 & 1,4\\ \hline
\end{tabular} 
\caption{Utilities of senders}  \label{tb8}
\end{subtable}
\vspace{-12pt}
\end{table}
\end{ex}
In this example, we demonstrate that all the conditions in Proposition \ref{lem:prelim} are either violated or not needed. First, although there exists a (unique) collaborative state $\StateX{TR}$ satisfying condition 1, \Sender{1}'s and \Sender{2}'s preference order on states $\{\StateX{BL},\StateX{BC}\}$ are fully aligned, instead of a polar opposite preference required by Proposition \ref{lem:prelim}. Actually, persuading receiver to take action $\Act{TR}$ is best not only for \Sender{1} but also for \Sender{2}. Hence, \Sender{2} will not threaten $\Sender{1}$ on her signals suggesting $\Act{TR}$, and the signalling strategy on persuading $\Act{TR}$ will be independent of the commitment order. Second, although the second condition (a demand of collaboration) holds on \Sender{1}'s private information $I_1=B$, Proposition \ref{lem:prelim} provides no guidance on how senders will collaborate without an effective collaborative state for the construction of collaborative signalling strategies. Thus, the second condition is not helpful without the existence of collaborative state. Third, none of the preference orders of \Sender{1} and \Sender{2} are polar-opposite under \Sender{1}'s truthful report. Since all conditions are either defunct or violated, Proposition \ref{lem:prelim} cannot help determine whether the commitment order matters or not in this example.

In actuality though, the commitment order does matter here. Note that \Sender{1} prefers a mixture of her private information $M,B$ to persuade the receiver in taking action $\Act{MC}$, but \Sender{2} prefers a separation of \Sender{1}'s private information $M$ and $B$. This gives \Sender{2} a chance to pose a credible threat. Note that $\StateX{MC}$ is not a collaborative state because \Sender{2} will first persuade the receiver to take \Act{MR}, and then help \Sender{1} persuade the receiver to take \Act{MC} when she cannot increase the value of  $\Prob{\Act{MR}}$. To verify that the commitment order matters in this example, we calculate the utilities of the senders \Sender{1}, \Sender{2} and the receiver under both commitment orders. For simplicity of representation, let $\Signal{1}^X$ denote \Sender{1}'s signal suggesting the action set $\{\Act{XL},\Act{XC},\Act{XR}\}$, $X\in\{T,M,B\}$ and $\Signal{2}^Y$ denote \Sender{2}'s signal suggesting the action set $\{\Act{TY},\Act{MY},\Act{BY}\}$, $Y\in\{L,C,R\}$. In other words, $\Signal{1}^X$ implies $\Prob{\StateX{XY}}=\max\{ \Prob{\StateX{TY}},\Prob{\StateX{MY}},\Prob{\StateX{BY}}\}$ for every $Y\in\{L,C,R\}$ and $\Signal{2}^Y$ implies $\Prob{\StateX{XY}}=\max\{ \Prob{\StateX{XL}},\Prob{\StateX{XC}},\Prob{\StateX{XR}}\}$ for every $X\in\{T,M,B\}$. We will use $\Signal{1}^X$ and $\Signal{2}^Y$ to state the commitments in this example.

\subsubsection{\texorpdfstring{\Sender{1}}{Sender 1} commits first}
First, both \Sender{1} and \Sender{2} wants to maximize the probability of $\Act{TR}$ according to their utility functions. Hence, \Sender{1} will first maximize the probability of signal $\Signal{1}^T$.
\begin{itemize}
    \item Send $\Signal{1}^T$ with probability $1$ when her private information $I_1=T$.
    \item Send $\Signal{1}^T$ with probability $0.25$ when her private information $I_1=M$.
    \item Send $\Signal{1}^T$ with probability $\tfrac{1}{7}$ when her private information $I_1=B$.
\end{itemize}

Now, conditional on \Sender{1} not sending signal $\Signal{1}^T$, the distribution of states reduces to Table~\ref{tb10}, where $w=0.735$ normalizes the total probability to be $1$.
\begin{table}[hbt]
\parbox{0.48\textwidth}{
\caption{Distribution of the states when $\Signal{1}\neq \Signal{1}^T$}
    \centering
  \begin{tabular}{|l|l|l|l|}  
\hline
  & L  & C & R   \\ \hline
T & $0$ & $0$ & $0$ \\ \hline
M & $0.09/w$ & $0.15/w$ & $0.015/w$ \\ \hline
B & $0.3/w$ & $0.18/w$ & $0$ \\ \hline
\end{tabular} \label{tb10}
}
\parbox{0.48\textwidth}{
\caption{Distribution of the states when  $\Signal{1}=\Signal{1}^M$}
    \centering
  \begin{tabular}{|l|l|l|l|}  
\hline
  & L  & C & R   \\ \hline
T & $0$ & $0$ & $0$ \\ \hline
M & $0.09/w$ & $0.15/w$ & $0.015/w$ \\ \hline
B & $0.25/w$ & $0.15/w$ & $0$ \\ \hline
\end{tabular} \label{tb18}
}
\end{table}

According to Table \ref{tb10}, \Sender{2} knows the true state is \StateX{MR} when her private information $I_2=R$, regardless of the signalling strategy \Sender{1} uses to mix $M$ and $B$. Since \Sender{2} prefers $\StateX{MR}$, then $\StateX{MC}$, and finally $\StateX{ML}$, \Sender{2} will first persuade the receiver on taking action $\Act{MR}$ when $\Signal{1}\neq \Signal{1}^T$. When \Sender{2} has maximized the total probability of $\Act{MR}$ taken by receiver conditional on her signalling strategies, \Sender{2} is willing to collaborate with \Sender{1} on persuading the receiver to take action $\Act{MC}$ instead of $\Act{ML}$. Therefore, \Sender{1}'s optimal mixture of information sets $M$ and $B$ is to maximize the conditional probability of $B$ in her signal, while ensuring that the posterior belief of $\StateX{MC}$ is no less than $\StateX{MB}$.
Using the concavification approach~\cite{kamenica2019bayesian}, the posterior beliefs of states conditional on \Sender{1} sending signal $\Signal{1}^M$ is presented in Table \ref{tb18}, where $w=0.655$ normalizes the total probability to be one.

\begin{figure}[ht]
\begin{subfigure}{0.49\textwidth}
    \centering
    \includegraphics[width=.95\textwidth]{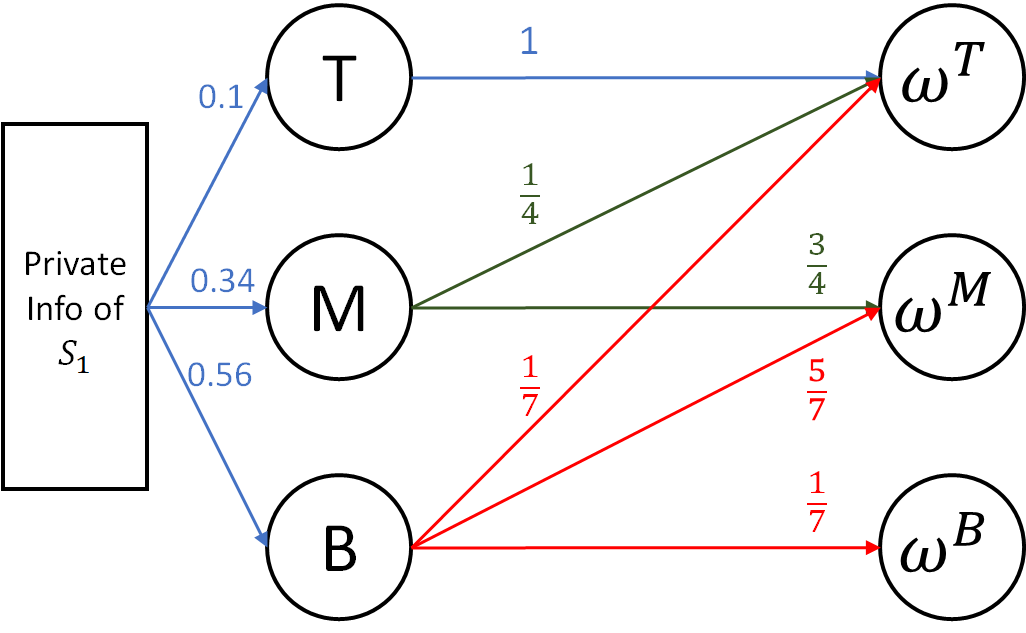}
    \subcaption{When \Sender{1} commits first}
    \label{fig:s1fcommit}    
\end{subfigure}
\begin{subfigure}{0.49\textwidth}
    \centering
    \includegraphics[width=.95\textwidth]{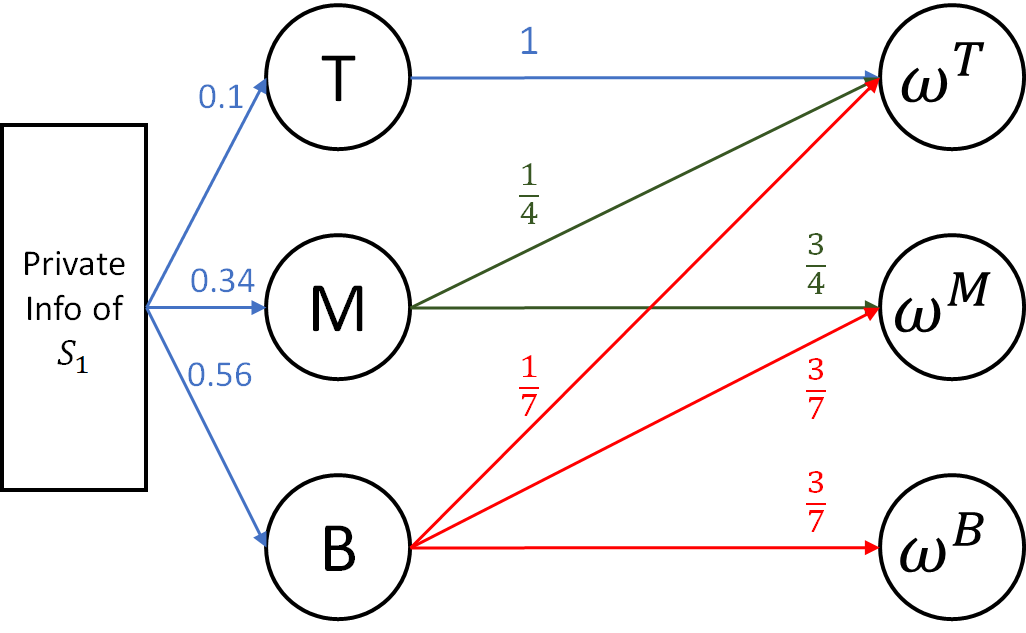}
    \subcaption{When \Sender{2} commits first}
    \label{fig:s1scommit}
\end{subfigure}
\vspace{-8pt}
\caption{Sender \Sender{1}'s optimal commitment}
\end{figure}

Given \Sender{1}'s signal which mixes information sets $T$, $M$, and $B$ and the optimal mixture of $M$ and $B$ derived above, \Sender{1} still has a positive probability on information set $B$ for which she has no choice but to tell the truth. By calculating the conditional probabilities of each state conditional on \Sender{1}'s private information and her optimal signalling strategy presented above, we can find the complete commitment of \Sender{1}. We depict it in Figure \ref{fig:s1fcommit}, where each number next to an arrow from \Sender{1}'s private signal $(T,M,B)$ to a signal $(\Signal{1}^T,\Signal{1}^M,\Signal{1}^B)$ represents the conditional probability of sending a particular signal under that private information.

Given \Sender{1}'s optimal commitment, deriving \Sender{2}'s optimal commitment is equivalent to solving a classical single-sender Bayesian persuasion problem for each of \Sender{1}'s realized signals when \Sender{2} commits after \Sender{1}. Thus, \Sender{2}'s optimal signalling strategy can be calculated in straightforward manner via the concavification approach that solves classical Bayesian persuasion problems. Since the result of concavification approach can be easily verified via calculating \Sender{2}'s utility change under small deviations, we omit the numerical computations of the concavification procedure and present \Sender{2}'s optimal commitment below:

\begin{itemize}
    \item While observing $\Signal{1}^T$, send $\Signal{2}^R$ with probability $1$, whatever her private information is.
    \item While observing $\Signal{1}^M$, send signals using the following strategy:
    \begin{itemize}
        \item Send $\Signal{2}^R$ with probability $1$ when her private information is $R$, with probability $0.1$ when her private information is $C$, and with probability $0.06$ when her private information is $L$.
        \item Send $\Signal{2}^C$ with probability $0.9$ when her private information is $C$ and with probability $0.54$ when her private information is $L$.
        \item Send $\Signal{2}^L$ with probability $0.4$ when her private information is $L$.
    \end{itemize}
    \item While observing $\Signal{1}^B$, send signals using the following strategy:
    \begin{itemize}
        \item Send $\Signal{2}^C$ with probability $1$ when her private information is $C$ and with probability $0.6$ when her private information is $L$.
        \item Send $\Signal{2}^L$ with probability $0.4$ when her private information is $L$.
    \end{itemize}
\end{itemize}

We remark that \Sender{1} plays an aggressive strategy which relies on \Sender{2}'s collaboration to persuade the receiver towards taking \Sender{1}'s preferred action. When \Sender{1} sends $\Signal{1}^M$, the interim belief of the state distribution is Table \ref{tb18} presented earlier:

We notice that state $\StateX{BL}$ has the highest probability if \Sender{2} stays silent. Without the signal of \Sender{2}, the receiver will take action $\Act{BL}$ and \Sender{1} experiences the lowest utility. However, since $\Act{MC}$ gives \Sender{2} a higher utility than $\Act{ML}$, \Sender{2} will collaborate with \Sender{1} after she maximizes the total probability of $\Act{MR}$. This allows \Sender{1} to commit to the above signalling strategy. Last, we calculate (detailed in Appendix \ref{app:utilcal4.1.1}) the expected utility of agents and obtain $ [\Expect{\PRIOR}{\UTIL{\Sender{1}}},\Expect{\PRIOR}{\UTIL{\Sender{2}}},\Expect{\PRIOR}{\UTIL{R}}]=[2.4336,2.6884,0.286]$.

\subsubsection{\texorpdfstring{\Sender{2}}{Sender 2} commits first}
Again, both \Sender{1} and \Sender{2} aim to maximize the probability of $\Act{TR}$ according to their utility functions. Hence, when \Sender{1} maximizes the probability of signal $\Signal{1}^T$, \Sender{2} has no incentive to restrict her signal $\Signal{1}^T$. Therefore, before \Sender{2} makes her commitment, she knows \Sender{1}'s signalling strategy on $\Signal{1}^T$ below.
\begin{itemize}
    \item Send $\Signal{1}^T$ with probability $1$ when her private information $I_1=T$.
    \item Send $\Signal{1}^T$ with probability $0.25$ when her private information $I_1=M$.
    \item Send $\Signal{1}^T$ with probability $\tfrac{1}{7}$ when her private information $I_1=B$.
\end{itemize}

Now, conditional on \Sender{1} not sending signal $\Signal{1}^T$, the distribution of states reduces to Table \ref{tb10} presented earlier. \Sender{2}'s problem then reduces to designing an optimal commitment under Table \ref{tb10}. Opposite to \Sender{1}, \Sender{2} prefers a separation of \Sender{1}'s information sets $M$ and $B$. Therefore \Sender{2} will restrict \Sender{1}'s mixture of $M$ and $B$ in her signal $\Signal{1}^M$ as much as she can by posing a credible threat. Here we detail \Sender{2}'s thought process on designing her threat strategy.

Since \Sender{1} will never use a signal $\Signal{1}$ which mixes $M$ and $B$ such that $\Prob{MC|\Signal{1}}\leq \Prob{BC|\Signal{1}}$\footnote{Otherwise, the receiver will never take action $\Act{MC}$ suggested by \Sender{1}.}, \Sender{2} can only restrict \Sender{1}'s mixture by tailoring her signal on her private signal $L$.
Ideally, \Sender{2} has two choices: (1) increase the probability of signal $\Signal{2}^R$ while receiving $\Signal{1}^M$ with private signal $L$, or (2) increase the probability of signal $\Signal{2}^C$ while receiving $\Signal{1}^M$ with private signal $L$. However, the first choice cannot threaten \Sender{1} since \Sender{1} has utility $0$ under $\Act{MR}$. This is because \Sender{1} is better off when \Sender{2} fails to persuade the receiver to take action $\Act{MR}$\footnote{In this case, the receiver will take action \Act{ML} instead of \Act{MR}, and \Sender{1} obtains a higher utility.}. Thus, \Sender{2} can only pick the second choice and tailor the probability of signal $\Signal{2}^C$ to restrict \Sender{1}. Given the inequality $\mathbb{P}_{\PRIOR}(\StateX{BL})>\mathbb{P}_{\PRIOR}(\StateX{BC})$ from the prior, \Sender{2} can reduce the probability $\Prob{\Signal{2}^L|\Signal{1}^M}$ to $0$ to restrict \Sender{1}'s mixture of $M$ and $B$. In other words, \Sender{1} acknowledges that the receiver will take either $\Act{MR}$ or $\Act{MC}$ when $\Signal{1}^M$ is sent. Given the interim distribution of Table \ref{tb10}, \Sender{2}'s optimal signalling scheme (after using the concavification approach to derive $\Signal{2}^R$) is to send signal $\Signal{2}^R$ with probability $0.1$ and $\Signal{2}^C$ with probability $0.9$ when she observes $\Signal{1}^M$ and her private information is $L$. Last, when \Sender{2} observes $\Signal{1}^B$, she immediately knows \Sender{1}'s private signal is $B$ (because mixing $T$ or $M$ to suggest either $\Act{BL}$ or $\Act{BM}$ hurts \Sender{1}). Thus, \Sender{2}'s optimal commitment can be solved via a concavification approach analogous to the one-sender, binary-state Bayesian persuasion problem. To sum up, we list \Sender{2}'s optimal signalling strategy below:

\begin{itemize}
    \item While observing $\Signal{1}^T$, send $\Signal{2}^R$ with probability $1$, whatever her private information is.
    \item While observing $\Signal{1}^M$, send signals using the following strategy:
    \begin{itemize}
        \item Send $\Signal{2}^R$ with probability $1$ when her private information is $R$, with probability $0.1$ when her private information is $C$, and with probability $0.1$ when her private information is $L$.
        \item Send $\Signal{2}^C$ with probability $0.9$ when her private information is $C$ or $L$.
    \end{itemize}
    \item While observing $\Signal{1}^B$, send signals using the following strategy:
    \begin{itemize}
        \item Send $\Signal{2}^C$ with probability $1$ when her private information is $C$ and with probability $0.6$ when her private information is $L$.
        \item Send $\Signal{2}^L$ with probability $0.4$ when her private information is $L$.
    \end{itemize}
\end{itemize}

Based on \Sender{2}'s commitment, \Sender{1} knows she cannot commit to an aggressive signalling strategy and wish \Sender{2} to help her persuading the receiver to take action
$\Act{MC}$. Moreover, her maximum mixture of $M$ and $B$ is restricted by \Sender{2} with the satisfaction of the inequality $\Prob{\StateX{BL}|\Signal{1}^M}\leq \Prob{\StateX{MC}|\Signal{1}^M}$. Thus, given the distribution presented in Table \ref{tb10}, \Sender{1} can only send signal $\Signal{1}^M$ with probability $\frac{1}{2}$ when her private information is $B$ and signal $\Signal{1}^T$ is not sent. In short, \Sender{1}'s optimal commitment is to first maximize $\Signal{1}^T$, and then maximize $\Signal{1}^M$ under the restriction of \Sender{2}'s signalling strategy. After calculating the probability of $\Signal{1}^T$ and the conditional probability of $\Signal{1}^M$, we summarize \Sender{1}'s optimal commitment below and a diagram in Figure \ref{fig:s1scommit}.
By calculating the expected utility of agents (detailed in Appendix \ref{app:utilcal4.1.2}), we obtain
$[\Expect{\PRIOR}{\UTIL{\Sender{1}}},\Expect{\PRIOR}{\UTIL{\Sender{2}}},\Expect{\PRIOR}{\UTIL{R}}]=[2.207,3.158,0.35]$.

\subsubsection{Comparison of utilities}        Similar to the example in Section~\ref{sec:4.3.1}, both senders want to commit first, and the receiver has a preference on letting \Sender{2} commit first in this example. However, the unique collaborative state in this example violates conditions in Proposition \ref{lem:prelim} and the signalling strategies (the probability of suggesting $\Act{TR}$) are the same in both commitment orders. However, the commitment order matters because of the collaboration in persuading $\Act{MC}$, even though $\Act{MC}$ is not \Sender{2}'s most preferred action under $I_1=M$.
Moreover, if senders can trade their commitment order and \Sender{1} is initially endowed the first commitment slot, every transfer $x\in(0.2266,0.4616)$ from \Sender{2} to \Sender{1} makes both senders and the receiver better off by trading the first commitment slot from \Sender{1} to \Sender{2}. If the transfer from the receiver to \Sender{1} is also allowed (even though the receiver doesn't participate in this trade), the maximum compensation $\Sender{1}$ could receive from giving out the first commitment slot grows to $0.5276$. This example demonstrates a case where both senders' and the receiver's utilities can increase when the first commitment slot is transferred from \Sender{1} to \Sender{2}. 

\subsection{General result: Sufficient conditions}
With the example above, we learned that the commitment order matters in a more general scenarios than those prsented in Proposition \ref{lem:prelim}. The analysis in Example \ref{ex:complex} suggests that when both a credible threat and a collaboration exist, the commitment order will matter. In order to know whether a (partial) collaboration is possible, we start by defining the conditional distribution of \Sender{1}'s information sets and use it to represent the best response signalling strategy of \Sender{2} under an observed signal of \Sender{1}.

\begin{dfn}
Given a prior $p$ and \Sender{1}'s commitment $\Gamma_1$, a signal realization $\Signal{1}$ indicates a conditional distribution of \Sender{1}'s information sets, $\Delta_{\Gamma_1}(\Infoset{1}|\Signal{1},p)$, which we denote as $q^{\Signal{1}}$.
\end{dfn}
When two signals\footnote{Two signals need not be from the same commitment.} $\Signal{1},\Signal{1}'$ have the same information set distribution, i.e., $q^{\Signal{1}}=q^{\Signal{1}'}$, the
best response of the receiver will be identical for these two signals. Thus, to avoid duplication and for simplicity of comparing different signals, we abuse notation and use $a^*(q^{\Signal{1}},\Signal{2})$ to represent the receiver's best response under given \Sender{1}'s information set distribution and $\Signal{2}$, i.e., $a^*(\Signal{1},\Signal{2}),a^*(\Signal{1}',\Signal{2})$. Since the receiver's utility is a function of \Sender{1}'s information set distribution and \Sender{2}'s signal, we are ready to define \Sender{2}'s best response signalling strategies.

\begin{dfn}
Given a prior $p$ and a distribution of $S_1$'s information set $q^{\Signal{1}}$, $G(p,q^{\Signal{1}})$ represents the set of best response signalling strategies of $S_2$ while observing $S_1$'s information set distribution $q^{\Signal{1}}$. 
\end{dfn}
Mathematically, an element $g\in G(p,q^{\Signal{1}})$ is a function that maps $I_2\in\Infoset{2}$ to a signal distribution $r\in \Delta(\SignalSpace_2)$ such that $\Expect{p,g}{U_{S_2}|q^{\Signal{1}},r}\geq \Expect{p,g'}{U_{S_2}|q^{\Signal{1}},r}$ for all $g' \not\in G(p,q^{\Signal{1}})$. For simplicity of representation, $G(p, q^{\Signal{1}}+\alpha q^{\bar{\omega}_1})$ denotes \Sender{2}'s best response under \Sender{1}'s mixed signal $\hat{\omega}_1$ such that $q^{\hat{\omega}_1}=\frac{1}{1+\alpha}q^{\Signal{1}}+\frac{\alpha}{1+\alpha}q^{\bar{\omega}_1}$. Since \Sender{2}'s expected utility is maximized in every best response strategy $g\in G(p,q^{\Signal{1}})$, i.e., $ \Expect{p}{U_{S_2}|\Signal{1},g}=\Expect{p}{U_{S_2}|\Signal{1},g'}$ for all $g,g'\in G(p,q^{\Signal{1}})$,
we again abuse notation and use $\Expect{p}{U_{S_2}|\Signal{1},G(p,q^{\Signal{1}})}$ to represent \Sender{2}'s expected utility under \Sender{1}'s information set distribution $q^{\Signal{1}}$, prior $p$, and \Sender{1}'s signal $\Signal{1}$.

Given the above definitions, we will present a set of sufficient conditions of when the commitment order matters that relies on two critical insights gained from Example \ref{ex:complex}:
\begin{enumerate}
    \item Both senders are willing to (partially) collaborate on at least a particular private information state of \Sender{1}.
    \item On such a ``collaborative" information state, \Sender{1} and \Sender{2} have a conflict of interest on the level of collaboration, which allows \Sender{2} to pose a credible threat.
\end{enumerate}

\begin{thm} \label{lem:suffcon}
Given a prior $p$, the commitment order matters if there exists a pair of $S_1$'s information sets $\Info{1}{x},\Info{1}{y}$ satisfying the following conditions:
\begin{enumerate}
    \item There exist two parameters $\alpha>\beta>0$ and a signalling strategy $\hat{\Gamma}_2$ satisfying the following conditions:
    \begin{enumerate}
        \item $G(p,\Info{1}{x}+\alpha \Info{1}{y})=G(p,\Info{1}{x}+\beta \Info{1}{y})$,
        \item Let $\Signal{1}^\alpha,\Signal{1}^\beta$ be two mock signals of \Sender{1} such that $\Prob{\Info{1}{x}|\Signal{1}^\alpha}=\frac{1}{1+\alpha}$,
    $\Prob{\Info{1}{y}|\Signal{1}^\alpha}=\frac{\alpha}{1+\alpha}$,
    $\Prob{\Info{1}{x}|\Signal{1}^\beta}=\frac{1}{1+\beta}$, and 
    $\Prob{\Info{1}{y}|\Signal{1}^\beta}=\frac{\beta}{1+\beta}$, then
    \begin{align*}
        \hat{\Gamma}_2(p,\Signal{1}^\beta)\in G(p,\Info{1}{x}+\beta \Info{1}{y})~\text{and}~\hat{\Gamma}_2(p,\Signal{1}^\alpha)\notin G(p,\Info{1}{x}+\alpha \Info{1}{y}),
    \end{align*}
        
        \item $\Expect{p}{U_{S_1}|\Signal{1}^\alpha,\hat{\Gamma}_2}<  \frac{\Prob{\Info{1}{x}}+\beta \Prob{\Info{1}{y}}}{\Prob{\Info{1}{x}}+\alpha \Prob{\Info{1}{y}}}\Expect{p}{U_{S_1}|\Signal{1}^\beta,\hat{\Gamma}_2}
        +\frac{(\alpha-\beta)\Prob{\Info{1}{y}}}{\Prob{\Info{1}{x}}+\alpha \Prob{\Info{1}{y}}}\Expect{p}{U_{S_1}|\Info{1}{y},\hat{\Gamma}_2}$,
        \item $\Expect{p}{U_{S_2}|\Signal{1}^\beta,G(p,\Info{1}{x}+\beta \Info{1}{y})}< \Expect{p}{U_{S_2}|\Info{1}{y},G(p, \Info{1}{y})}$.
    \end{enumerate}
    \item Sender $S_1$ has a higher expected utility under $\Info{1}{x}$ than $\Info{1}{y}$, regardless of \Sender{2}'s tie-breaking rule, i.e.,
    \begin{align*}
    \min_{g\in G(p,\Info{1}{x})}\Expect{p,g}{U_{S_1}|\Info{1}{x}}> \max_{g'\in G(p,\Info{1}{y})}\Expect{p,g'}{U_{S_1}| \Info{1}{y}}.        
    \end{align*}

    \item There is no commitment pair $(\Gamma_1,\Gamma_2)$ such that
    \begin{align*}
        \sum_{\StateX{i}\in \Info{1}{y}}\mathbb{P}_{p,\Gamma_1,\Gamma_2}(\Act{i})=0 \text{~or~}\sum_{\StateX{i}\in \Info{1}{x}}\mathbb{P}_{p,\Gamma_1,\Gamma_2}(\Act{i})=0
    \end{align*}
\end{enumerate}
\end{thm}

Although the conditions in Theorem \ref{lem:suffcon} look technical, there are intuitions behind each condition. The first condition in Theorem \ref{lem:suffcon} states that the sender $S_2$ prefers a separation of sender $S_1$'s information sets $\Info{1}{x}$ and $\Info{1}{y}$. Moreover, if \Sender{2} can commit first, she can make a credible threat on her commitment to constrain the mixing of $\Info{1}{x}$ and $\Info{1}{y}$ in sender $S_1$'s signalling strategies. This generalizes condition 3 in Proposition \ref{lem:prelim}. Besides, the combination of condition 1(a) and 1(d) guarantees that \Sender{2} can collaborate with \Sender{1}, and this generalizes the collaborative state requirement in Proposition \ref{lem:prelim}. The second condition in Theorem \ref{lem:suffcon} states that the sender $S_1$ prefers a mixture of $\Info{1}{x}$ and $\Info{1}{y}$ in her signalling strategy. Combining the intuition behind the first two conditions, the two senders have opposite preferences on the mixture of $\Info{1}{x}$ and $\Info{1}{y}$. Based on the first two conditions, senders $S_1, S_2$ are already in a scenario where the sender $S_2$ is capable of constraining $S_1$'s mixture of $\Info{1}{x}$ and $\Info{1}{y}$, and a collaboration between \Sender{1} and \Sender{2} is possible under $\Info{1}{x}$. Hence, the proof of Theorem \ref{lem:suffcon} shares the same general principle with the proof of Proposition \ref{lem:prelim}, i.e., constructing a credible threat using an information set where both sender collaborate on persuasion. The third condition is a regularity condition which guarantees that the prior $\PRIOR$ has put enough probability mass in states $\theta \in \Info{1}{x}$ and $\theta \in \Info{1}{y}$. This is to avoid corner cases where the senders $S_1$ and $S_2$ have opposite preferences on the mixture of $\Info{1}{x}$ and $\Info{1}{y}$, but their opposite preferences do not matter because the sender $S_1$ can commit to some signalling strategies which never suggests an action $\Act{i}\in \{\Act{j}|\StateX{j} \in \Info{1}{x}\cup \Info{1}{y}\}$. The statement of condition 3 may demand a search of the whole space of possible commitments, which could make this theorem difficult to apply in models with a large state space. However, in most practical problems, verifying condition 3 can be done in straightforward manner by inspecting the prior distribution. Hence, we believe condition 3 doesn't undermine the contribution of Theorem \ref{lem:suffcon}.

\subsection{General result: Necessary conditions}

The first two conditions in Theorem \ref{lem:suffcon} capture the scenario when a credible threat by $S_2$ can be issued, hence constraining $S_1$'s signalling strategies. This leads to the question of whether there are simpler and more basic requirements that make credible threats possible without analyzing the best response of commitments. Moreover, results in Theorem \ref{lem:suffcon} and Proposition \ref{lem:prelim} depend on the prior. Hence, we would also like to obtain more basic conditions that are prior independent. Last, we observe that different tie-breaking rules of the receiver may play a role in determining whether the commitment order matters. However, it is hard to determine the criticality of the receiver’s tie-breaking rule under sufficient conditions because each tie-breaking rule only plays a role under a small set of priors. With these considerations in mind, we present in the following theorem a set of necessary conditions of scenarios where the commitment order could matter.

\begin{thm} \label{lem:necessary}
Given a prior $p$ and both senders' utility functions $U_{S_1}$ and $U_{S_2}$, if both conditions below are violated, then the commitment order does not matter:

\begin{enumerate}
    \item There exists a set of states $(\StateX{\alpha}, \StateX{\beta}, \StateX{\gamma})$ satisfying the following properties:
    \begin{enumerate}
    \item $\theta_{\alpha}$ and $\theta_{\beta}$ are in the same information set of \Sender{1} but $\StateX{\gamma}$ is not, i.e.,$\exists! \Info{1}{x}\ni \theta_{\alpha}, \theta_{\beta}$ and $\theta_{\gamma} \in \Info{1}{y} \neq \Info{1}{x}$. Besides, $\theta_{\alpha}$ and $\theta_{\beta}$ are not in the same information set of \Sender{2}, i.e., $\exists~ \Info{2}{x}\neq \Info{2}{y}$ s.t. $\theta_{\alpha} \in \Info{2}{x}$ and $\theta_{\beta} \in \Info{2}{y}$. But $\StateX{\gamma}$ is in the same information set as either $\StateX{\alpha}$ or $\StateX{\beta}$, i.e.,
    $\theta_{\gamma} \in \Info{2}{x}$ or $\theta_{\gamma} \in \Info{2}{y}$. An example relationship is depicted in Figure \ref{fig:necefig}.
    \begin{figure}[ht]
    \centering
    \includegraphics[width=0.4\textwidth]{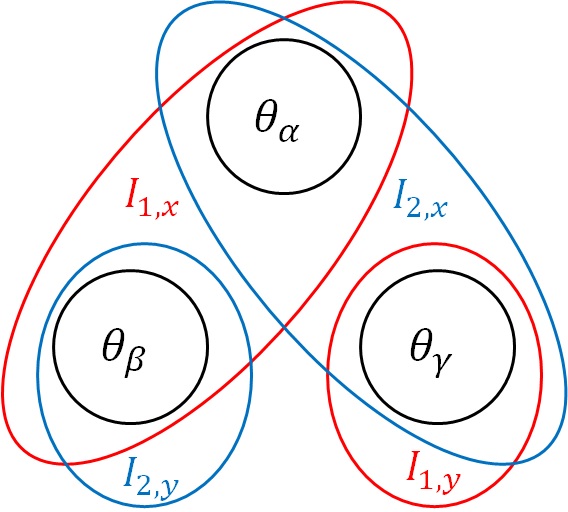}
    \caption{An example information structure of a state tuple in condition (a)}
    \label{fig:necefig}
\end{figure}
    \item The utilities of \Sender{1} and \Sender{2} satisfy
    \begin{align}
        &U_{S_1}(\Act{\alpha})>\max \{U_{S_1}(\Act{\beta}),U_{S_1}(\Act{\gamma})\}, \\
        &U_{S_2}(\Act{\gamma})>U_{S_2}(\Act{\alpha})>U_{S_2}(\Act{\beta}).
    \end{align}
\end{enumerate}
\item The receiver's tie-breaking rule is belief dependent.
\end{enumerate}

\end{thm}

The first condition in Theorem~\ref{lem:necessary} captures a requirement of an alignment between \Sender{1}'s and \Sender{2}'s utilities. The requirement is considerably weaker than the definition of collaborative states and condition 1 of Theorem \ref{lem:suffcon}. The commitment order may matter when both senders want to collaborate in a pair of states, but they have a conflict of interest on mixing a third state\footnote{This third state is in a different information set of \Sender{1}.} with this pair. When both senders have no conflicts of interest or never collaborate, then the commitment order will not matter. To avoid confusion, we note here that the violation of Theorem \ref{lem:necessary} does not demand fully-aligned preference orderings between \Sender{1} and \Sender{2}, or polar opposite preference orderings between \Sender{1} and \Sender{2}. Senders can still collaborate in a set of states and have a conflict of interest in another set of states, as long as these two sets do not intersect. The second condition states that if the receiver's tie-breaking rule depends on her beliefs, then a sender may tailor her signalling strategies to make the tie-breaking rule favor her instead of the other sender. When this occurs, the sender committing first suffers due to the tie-breaking rule. The sender committing later can always tailor the receiver's beliefs on top of the earlier sender's commitment and make the tie-breaking rule favor her instead. 

\section{Discussion} \label{sec:commitdis}

We present two examples in this discussion section to demonstrate interesting phenomena. The first example demonstrates an interesting scenario where the commitment order matters. In this scenario, the only reason that drives \Sender{1} to get the first commitment slot is because she can stay silent while committing first. The second example compares sequential commitments with simultaneous commitments. In the second example, the sequential commitment setting gives the receiver higher expected utility and more informative signals than simultaneous commitments. Beside these two examples, we remark on the difficulty of developing general algorithms to derive the optimal commitments under incomplete information of senders to close the discussion section.

\subsection{Silence is golden: commit to send a non-informative signal}
We present an interesting example where commitment order matters, but senders commit completely opposite signalling strategies while committing first. When \Sender{1} commits first, her optimal signalling strategy is either to stay silent or to send non-informative signals. However, when \Sender{2} commits first, her optimal commitment is a truth-telling strategy, and her truth-telling strategy forces \Sender{1}'s optimal commitment to become truth-telling as well. Hence, in the example detailed below, if the receiver can choose the commitment order, she prefers \Sender{2} committing first. 
\begin{ex}
    The state space is ternary $\StateSpace=\{\StateX{1}, \StateX{2}, \StateX{3}\}$ with prior distribution $(\Prob{\StateX{1}},\Prob{\StateX{2}},\Prob{\StateX{3}})=\big(0.2,0.5,0.3\big)$. \Sender{1} knows the state is $\StateX{3}$ or not, but cannot distinguish $\StateX{1}$ and $\StateX{2}$, $\Infoset{1}=\{\{\StateX{1},\StateX{2}\},~\{\StateX{3}\}\}$. \Sender{2} knows the state is $\StateX{1}$ or not, but cannot distinguish $\StateX{2}$ and $\StateX{3}$, $\Infoset{2}=\{\{\StateX{1}\}~,\{\StateX{2},\StateX{3}\}\}$. \Sender{1} and \Sender{2}'s utilities only depend on the receiver's action and are given by 
    \begin{align*}
        [U_{\Sender{1}}(\Act{1}), U_{\Sender{1}}(\Act{2}), U_{\Sender{1}}(\Act{3})]=[3,0,1],\quad [U_{\Sender{2}}(\Act{1}), U_{\Sender{2}}(\Act{2}), U_{\Sender{2}}(\Act{3})]=[1,0,3].
    \end{align*}
The receiver gets utility $1$ if the index of the state matches the action, and $0$ otherwise.
\end{ex}    
When \Sender{1} commits first, she knows that \Sender{2} cannot distinguish $\StateX{2}$ and \StateX{3}. Hence, as long as the receiver will guess on $\Act{2}$ as opposed to $\Act{3}$ given the posterior belief of the states, \Sender{2} is willing to collaborate with \Sender{1} to suggest \Act{1}. Hence, to maximize \Sender{1}'s expected utility, her optimal commitment is to stay silent, i.e., commit to a non-informative signal because $\Prob{\StateX{2}}>\Prob{\StateX{3}}$ in the prior. Given \Sender{1}'s non-informative commitment, \Sender{2}'s optimal commitment is to suggest $\Act{1}$ while she knows the state is $\StateX{1}$, and suggest $\Act{1}$ with probability $0.4$ when the state is either $\StateX{2}$ or $\StateX{3}$ (since \Sender{2} cannot distinguish between $\StateX{2}$ or $\StateX{3}$).

When \Sender{2} commits first, she knows \Sender{1} is willing to help her to persuade the receiver to take $\Act{2}$ when the information set contains only \StateX{2} and \StateX{3}. Hence, her optimal commitment is, to tell the truth, sending a signal to reveal $\StateX{1}$ when the true state is $\StateX{1}$, and sending a signal to tell she knows the state is $\StateX{2}$ or $\StateX{3}$ otherwise. Given \Sender{2}'s truth-telling commitment and underlying prior, \Sender{1} knows if she doesn't reveal $\StateX{2}$ when the true state is $\StateX{2}$, the receiver will take $\Act{3}$ when the state is $\StateX{2}$. Hence, her optimal signalling strategy becomes a truth-telling strategy as well.

\subsection{Comparison to simultaneous commitments}
When senders only have partial information (of the state of the world), sequential commitments may bring more informative signals jointly for the receiver. This is different from the result in Li-Norman~\cite{li2018sequential}, where senders have complete information. Hence, we want to compare our example in Section \ref{sec:4.3.1} with simultaneous commitments (and simultaneous signalling) from the receiver's perspective. In order to make a fair comparison, the prior and sender's utilities are the same as the example in Section \ref{sec:4.3.1}.  

\begin{table}[ht]
 \caption{State distribution and senders' utilities}
   \label{tb:example_simultaneous_commitment}
 \vspace{-8pt}
  \begin{subtable}[t]{0.48\textwidth}
  \centering
\begin{tabular}{|l|l|l|}  
\hline
  & L   & R   \\ \hline
T & $0.1$ & $0.2$ \\ \hline
B & $0.4$ & $0.3$ \\ \hline
\end{tabular} 
    \caption{Distribution of the states}
	\label{tb11}
  \end{subtable}
  \begin{subtable}[t]{0.48\textwidth}
  \centering
\begin{tabular}{|l|l|l|l|l|} 
\hline
  & $\Act{TL}$   &   $\Act{TR}$ &   $\Act{BL}$ &   $\Act{BR}$\\ \hline
$(U_{S_1},U_{S_2})$ & 1,0 & 2,2 & 0,0 & 0,3 \\ \hline
\end{tabular} 
 \caption{Utilities of senders}
	\label{tb9}
\end{subtable}
\vspace{-12pt}
\end{table}

From \Sender{2}'s perspective, since $TR$ is better than $TL$ and $BR$ is better than $BL$ based on her utility function, her commitment will try to persuade the receiver in the best possible manner for taking action $TR$ or $BR$, regardless of what \Sender{1} sends. Hence, \Sender{2}'s optimal commitment is the following:
\begin{itemize}
\item When her private information is $R$, send a signal to suggest $R$ w.p. 1.
\item When her private information is $L$, send a signal to suggest $R$ with probability $\frac{3}{4}$ and send a suggestion $L$ with probability $\frac{1}{4}$.
\end{itemize}

From \Sender{1}'s perspective, even though she cannot observe \Sender{2}'s commitment and signals, she can anticipate \Sender{2}'s optimal commitment via \Sender{2}'s utility function and the prior. Hence, although signals are sent and commitments are made simultaneously, \Sender{1} can use her inference on \Sender{2}'s optimal commitment to make her commitment. Thus, \Sender{1}'s commitment is the following:
\begin{itemize}
\item When her private information is $T$, send a signal to suggest $T$ w.p. 1.
\item When her private information is $B$, send a signal to suggest $T$ with probability $\frac{2}{3}$ and send a suggestion $B$ with probability $\frac{1}{3}$.
\end{itemize}

Given the above commitments, the receiver's utility is calculated below:
\begin{equation*}
    \Expect{\PRIOR}{\UTIL{R}}= 0.1\times 0+ 0.2 \times 1+0.4\times \tfrac{1}{3}\times\tfrac{1}{4}+0.3 \times \tfrac{1}{3}=\frac{1}{3}.
\end{equation*}
In Section \ref{sec:4.3.1}, the receiver's utility is $\frac{1}{3}$ while \Sender{1} commits first and $0.4$ while \Sender{2} commits first. In this example, simultaneous commitment (and signalling) grants the receiver the same utility as the case where \Sender{1} commits first. However, it is still a lower utility compared to the case where \Sender{2} commits first. We note that even though the commitments are made simultaneously, and signals are sent simultaneously, the equilibrium commitment pairs may not be unique when senders only have partial information about the state of the world. The non-uniqueness of equilibrium commitment pairs makes it difficult to make a fair comparison since equilibrium selection is also needed.

\subsection{Algorithmic challenges for deriving the optimal commitments} \label{sec:hardcommit}

Finally, we conclude by remarking that a general algorithm that solves the optimal commitments in information design problems with multiple senders where each sender obtains incomplete private information is challenging to develop. The reason is that the (required) level of inference depends on the information structure of the problem. Even when the commitment order is aligned with the signalling order, the backward-iteration algorithm presented in \cite{li2018sequential} cannot solve for the optimal commitments with partially informed senders. 
In general, deriving the optimal commitments demands iterative reasoning of the other sender's optimal signalling strategy under different interim beliefs of states. This process is straightforward when senders' information spaces are binary, but becomes cumbersome when senders' information spaces grow, even from binary to ternary as in Example \ref{ex:complex}. 
We believe deriving a general algorithm that iteratively reasons senders' optimal signalling strategies under incomplete information, will lead to further significant contributions to the field of information design, but this is beyond the scope of this work.



\bibliographystyle{ACM-Reference-Format}
\bibliography{X_Ref_incentivizinginfo}
\section{Appendix: An example justifying permutation-free commitments} \label{app:permutation_free}
There are four possible states of the world $\Theta=\{TL,TR,BL,BR\}$ (top-bottom and left-right) with prior distribution depicted in Table~\ref{tb5}. Sender \Sender{1} knows whether the state of the world is top or bottom $\Infoset{1}=\{T,B\}$ and sender \Sender{2} knows whether the state of the world is left or right $\Infoset{2}=\{L,R\}$. \Sender{1} can send signals $\Signal{1}\in \{\Signal{1,1},\Signal{1,2}\}$ and \Sender{2} can send signals $\Signal{2}\in \{\Signal{2,1},\Signal{2,2}\}$. We assume that both senders' utilities are as given in Table~\ref{tb4}, and this example requires sender $S_2$ to commit before \Sender{1}. All participants' utility functions and the (prior) state distribution are common knowledge. 
\begin{table}[ht]
 \caption{State distribution and senders' utilities}
  \label{tb:example_permutation_free}
 \vspace{-8pt}
  \begin{subtable}[t]{0.48\textwidth}
  \centering
\begin{tabular}{|l|l|l|}
\hline
  & L   & R   \\ \hline
T & $0.1$ & $0.2$ \\ \hline
B & $0.5$ & $0.2$ \\ \hline
\end{tabular} 
    \caption{Distribution of the states}
	\label{tb3}
  \end{subtable}
  \begin{subtable}[t]{0.48\textwidth}
  \centering
\begin{tabular}{|l|l|l|l|l|} 
\hline
  & $\Act{TL}$   &   $\Act{TR}$ &   $\Act{BL}$ &   $\Act{BR}$\\ \hline
$(U_{S_1},U_{S_2})$ & 2,0 & 4,2 & 3,0 & 0,3 \\ \hline
\end{tabular}
 \caption{Utilities of senders}
	\label{tb4}
\end{subtable}
\end{table}

Given the prior and the utility functions, sender \Sender{1} and \Sender{2}'s objective functions are the following: 
\begin{align*}
    \Gamma_1^{s*}&\in\arg\max_{\Gamma_1^s\in\mathbf{\Gamma_1}}\Expect{p}{U_{S_1}|\Gamma_2^s}\\ &\text{~s.t.~}  \Expect{p}{U_{S_2}|\Gamma_1^{s}(\Gamma_2^{s*})}\geq \Expect{p}{U_{S_2}|\Gamma_1^{s'}(\Gamma_2^{s^*})}, \\ &\text{~where~} \Gamma_1^{s'} \text{~is a commitment that swaps the signals of~} \Gamma_1^{s}, i.e.,\\  &\mathbb{P}_{\Gamma_1^{s}}(\Signal{1,1}|T)=\mathbb{P}_{\Gamma_1^{s'}}(\Signal{1,2}|T) \text{~and~} \mathbb{P}_{\Gamma_1^{s}}(\Signal{1,1}|B)=\mathbb{P}_{\Gamma_1^{s'}}(\Signal{1,2}|B).\\
    \Gamma_2^{f*}&\in\arg\max_{\Gamma_2^f\in\mathbf{\Gamma_2}}\Expect{p}{U_{S_2}|\Gamma_1^{s*}(\Gamma_2^s)}
\end{align*}
Since \Sender{1} cannot threaten \Sender{2} via signal reordering, \Sender{2} can commit to signalling strategies depending on the realization of \Sender{1}'s signal. For simplicity of the representation, let's assume $\Signal{1,1}$ has higher posterior belief on \Sender{1}'s private information $T$ than $\Signal{1,2}$. \Sender{2}'s optimal commitment is the following:

\begin{itemize}
\item While observing $\Signal{1,1}$, send $\Signal{2,1}$ with probability 1 no matter what private information she knows.
\item While observing $\Signal{1,2}$ and given her private information is $R$, send $\Signal{2,1}$ with probability 1.
\item While observing $\Signal{1,2}$ and given her private information is $L$, send signal $\Signal{2,1}$ with probability $0.4-\epsilon$ and $\Signal{2,2}$ with probability $0.6+\epsilon$, where $\epsilon$ is infinitesimal, i.e., $|\epsilon| \ll 1$.
\end{itemize}

Given \Sender{2}'s commitment above, the optimal commitment of \Sender{1} is the following:
\begin{itemize}

\item Given her private information $T$, send \Signal{1,1}.

\item Given her private information $B$, send \Signal{1,1} $40\%$ of the time and and send \Signal{1,2} $60\%$ of the time.
\end{itemize}
Given this pair of commitments, the expected utilities of all the agents are given by $(U_{S_1},U_{S_2},U_{R})=(2.86,1.88,0.5)$.

However, when \Sender{2} cannot commit to permutation-free commitments, \Sender{1}'s objective function changes to the following
\begin{align*}
\Gamma_1^{s*}\in\arg\max_{\Gamma_1^s\in\mathbf{\Gamma_1}}\Expect{p}{\UTIL{S_1}|\Gamma_2^s}
\end{align*}
If \Sender{2} commits to the same signalling strategy, \Sender{1} can simply swap \Signal{1,1} with \Signal{1,2} to get a higher expected utility. Under the threat of \Sender{1}'s reordering of signals, \Sender{2}'s commitment reduces to the following form.
No matter the signal sent by \Sender{1}, \Sender{2} commits to the following signalling scheme:
\begin{itemize}
    \item When \Sender{2}'s private information is $R$, send \Signal{2,1} with probability 1.
    \item When \Sender{2}'s private information is $L$, send signal \Signal{2,1} with probability $40\%-\epsilon$ of the time and and send \Signal{1,2} $60\%+\epsilon$ of the time, where $\epsilon$ is infinitesimal, i.e., $|\epsilon| \ll 1$.
\end{itemize}
Under \Sender{2}'s commitment above, \Sender{1}'s optimal commitment stays the same. However, the expected utility of agents changes to $(U_{S_1},U_{S_2},U_{R})=(2.74,1.76,0.56)$.

In this example, when permutation-free commitments are not enforced, \Sender{2} is under the threat of \Sender{1}'s reordering of signals, so her optimal strategy is to make her commitment independent of \Sender{1}'s signal realizations. Here, both senders suffer (utility losses) from the lack of permutation-free commitments. 
\section{Appendix: Proofs and Calculations}
\subsection{Proof of Proposition \ref{lem:prelim}} \label{pf:prop_prelim}
\begin{proof}
According to condition 2, we have a signal $\hat{\omega}_1$ which tells \Sender{1}'s private information is $I_1$, i.e., $\Prob{I_1|\hat{\Signal{1}}}=1$. First, we prove \Sender{1} does not prefer sending the signal $\hat{\omega}_1$. When \Sender{1} sends signal $\hat{\omega}_1$, \Sender{2} knows \Sender{1}'s private information is $I_1$. Let $\Gamma_2^*(I_1)$ denote \Sender{2}'s optimal signalling strategy while receiving $\hat{\omega}_1$, inequality (\ref{eq:4.9}) states \Sender{1}'s utility under action $\Act{\hat{\STATE}}$ is higher than her expected utility conditional on her signal $\hat{\omega}_1$. Thus, \Sender{1} prefers a mixture of her information set $I_1$ and the information set which contains the collaborative state $\hat{\STATE}$, called $\hat{I}_1$. 

We notice that if \Sender{2}'s optimal signalling strategy while receiving $\hat{\omega}_1$ is not unique, the conditional expectation $\Expect{p,\Gamma_2^*(I_1)}{\Util{\Sender{1}}{\STATE}{\ACT^*(I_1,\Signal{2})}|I_1}$ may not be well-defined since different \Sender{2}'s tie-breaking strategies may vary \Sender{1}'s expected utility. To deal with the non-uniqueness of $\Gamma_2^*(I_1)$, we assume the polar opposite preference ordering on the receiver's possible optimal action set given \Sender{1}'s private information $I_1$, called $\Phi(I_1)$ in condition 3. Therefore, when $\Sender{2}$ is indifferent between two different actions $\Act{m}$ and $\Act{n}$ where $\Act{m},\Act{n}\in \Phi(I_1)$, \Sender{1} is indifferent between \Act{m} and \Act{n} too. Thus, the conditional expectation $\Expect{p,\Gamma_2^*(I_1)}{\Util{\Sender{1}}{\STATE}{\ACT^*(I_1,\Signal{2})}|I_1}$ in inequality (\ref{eq:4.9}) is well-defined, and \Sender{1} prefers a mixture of $I_1$ and $\hat{I}_1$ when the inequality (\ref{eq:4.9}) holds.


Next, we prove \Sender{2} prefers a separation of $I_1$ and $\hat{I}_1$. Given assumption 4 in the problem formulation, where the true state can be revealed under both senders' truth-telling strategies, \Sender{2} knows the true state while observing the signal realization $\hat{\omega}_1$ and her private information $I_2$, and her optimal signalling strategy under $\hat{\omega}_1$ can be calculated directly via the concavification approach --- this is equivalent to solving classic Bayesian persuasion problem in \cite{kamenica2011bayesian} with finite states. According to the inequality \ref{eq:4.8} stated in condition 1 and the definition of collaborative state, for every $\StateX{k}\in \hat{I}_1$,the following inequality holds:
\begin{align*}
    \Util{\Sender{2}}{\StateX{k}}{\Act{\StateX{k}}}<        \Expect{p,\Gamma_2^*(I_1)}{\Util{\Sender{2}}{\STATE}{\ACT^*(I_1,\Signal{2})}|I_1}.
\end{align*}
Thus, \Sender{2} enjoys a higher expected utility while observing $\hat{\omega}_1$ than observing the signal telling her \Sender{1}'s private information is $\hat{I}_1$. Thus, \Sender{2} prefers a separation of $I_1$ and $\hat{I}_1$.

When condition 3 in Proposition \ref{lem:prelim} holds, i.e., preference orders of senders on $\Phi(I_1)$ are polar opposite, \Sender{2}'s expected-utility maximization strategy (via the concavification approach) will directly minimize \Sender{1} expected utility when $\hat{\Signal{1}}$ is realized. Therefore, from \Sender{1}'s objective, \Sender{1} should minimize the probability $\Prob{\hat{\Signal{1}}}$ to reduce the probability where \Sender{2} observes $\hat{\Signal{1}}$. Hence, $S_1$ will increase $\Prob{\Signal{1}|I_1}$ for every $\Signal{1}$ such that $\Expect{\PRIOR}{\UTIL{\Sender{1}}|\Signal{1}}>\Expect{\PRIOR}{\UTIL{\Sender{1}}|\hat{\Signal{1}}}$. This includes the signal which suggests the receiver to take the collaborative action $\Act{\hat{\STATE}}$. Hence, when \Sender{1} commits first, she will maximize the probability of suggesting the collaborative state $\hat{\STATE}$ via mixing $I_1$ with $\hat{I}_1$. Since $\hat{\STATE}$ is a collaborative state, \Sender{2} will collaborate with \Sender{1} while observing \Sender{1}'s signal which suggests $\Act{\hat{\STATE}}$. 

However, when sender \Sender{2} commits first, the inequality (\ref{eq:4.8}) stated in condition 1, i.e., $\Util{\Sender{2}}{\hat{\STATE}}{\Act{\hat{\STATE}}}<\Expect{p,\Gamma_2^*(I_1)}{\Util{\Sender{2}}{\STATE}{\ACT^*(I_1,\Signal{2})}|I_1}$, guarantees that \Sender{2} prefers the realized signal $\hat{\Signal{1}}$ instead of the realized signal which suggests her to collaborate toward $\hat{\STATE}$ via a mixture of $I_1$ and $\hat{I}_1$. Therefore, when sender \Sender{2} commits first, she will commit to a signalling strategy which minimizes the probability of $I_1$ mixed with $\hat{I}_1$ (in order to increase the total probability of observing $\hat{\Signal{1}}$). Since condition 2 guarantees that $\Prob{\hat{\omega}_1}>0$ for every optimal commitments, we will never fall into the corner case where \Sender{1} and \Sender{2} have an opposite preference on the mixture of $I_1$ and $\hat{I}_1$, but one of the sender's strategy of minimizing/maximizing the mixture of $I_1$ and $\hat{I}_1$ cannot be achieved owing to the low value of $\Prob{I_1}$ under the given prior. Hence, according to the opposite strategies of \Sender{1} and \Sender{2} on the mixture of $I_1$ and $\hat{I}_1$, the commitment order matters.
\end{proof}

\subsection{Proof of Claim \ref{clm:4}}
\begin{proof}
We prove this claim by contradiction. First, assume that both conditions are violated and the signal $\hat{\omega}_1$ revealing $I_1$ is the signal realization which gives \Sender{1} the highest expected utility under an equilibrium derived by solving the objectives stated in Section \ref{sec:objIV}, where each sender's commitment is an optimal one under the given commitment order. Then, let signal $\bar{\omega}_1$ represents the signal that $\Sender{1}$ sends when the true state is $\hat{\STATE}$, the violation of condition 2 guarantees that the following inequality holds
\begin{align*}
    \Expect{\PRIOR}{U_{S_1}(a^*_{\Gamma_1^*,\Gamma_2^*}(\hat{\omega}_1,\Signal{2}))}>\Expect{\PRIOR}{U_{S_1}(a^*_{\Gamma_1^*,\Gamma_2^*}(\bar{\omega}_1,\Signal{2}))},
\end{align*}
where $a^*_{\Gamma_1^*,\Gamma_2^*}(\Signal{1},\Signal{2})$ is the best response of the receiver under a realized pair of signal $\Signal{1},\Signal{2}$ and both senders' commitments $\Gamma_1^*$ and $\Gamma_2^*$. Since the signal $\hat{\omega}_1$ reveals the private signal $I_1$, we can construct another commitment $\Gamma_1'$ below, where $\epsilon \ll 1$:

\begin{enumerate}
    \item $\Prob{\hat{\omega}_1'}=\Prob{\hat{\omega}_1}+\epsilon \Prob{\bar{\omega}_1}$ \text{~and~} $\mathbb{P}(\theta|\hat{\omega}_1')=\frac{1}{1+\epsilon}\mathbb{P}(\theta|\hat{\omega}_1)+\frac{\epsilon}{1+\epsilon}\mathbb{P}(\theta|\bar{\omega}_1)$ for all $\theta\in \Theta$;
    \item $\mathbb{P}(\bar{\omega}_1')=(1-\epsilon) \mathbb{P}(\bar{\omega}_1)$, and the interim beliefs of $\bar{\omega}_1'$ and $\bar{\omega}_1$ are the same, i.e., $\mathbb{P}_{p,\Gamma_1'}\big(\STATE|\bar{\omega}_1'\big)=\mathbb{P}_{p,\Gamma_1}\big(\STATE|\bar{\omega}_1\big)$ for all $\STATE\in \StateSpace$;
    \item For every $\Signal{1}$ under the commitment $\Gamma_1'$ such that $\Signal{1} \neq \bar{\omega}_1$ and  $\Signal{1}\neq\hat{\omega}_1$,  $\mathbb{P}_{p,\Gamma_1'}(\Signal{1})=\mathbb{P}_{p,\Gamma_1^*}(\Signal{1})$ and
    $\mathbb{P}_{p,\Gamma_1'}\big(\STATE|\Signal{1}\big)=\mathbb{P}_{p,\Gamma_1^*}\big(\STATE|\Signal{1}\big)$ for all $\STATE\in \StateSpace$;
    
\end{enumerate}
As assumed earlier, $\hat{\omega}_1$ is the signal which gives $\Sender{1}$ the highest expected utility with $\mathbb{P}(I_1|\hat{\omega}_1)=1$, and $\bar{\omega}_1$ is the signal \Sender{1} sends to collaborate with \Sender{2} on persuading the receiver to take $\Act{\hat{\STATE}}$. Given the inequality (\ref{eq:4.9}), \Sender{2}'s optimal signalling strategy belongs to one of the following two cases:

\textbf{Case 1: \Sender{2} cannot elicit the collaborative state $\hat{\STATE}$ under the signal $\hat{\omega}_1'$.}\\
In this case, because $\epsilon\ll 1$, the optimal commitment of sender $S_2$ stays unchanged between $\Gamma_1^*$ and $\Gamma_1'$ since \Sender{2} cannot successfully persuade the receiver to take $\Act{\hat{\STATE}}$ and she has already chosen her optimal partition of $I_1$ in her commitment. Therefore, when $\epsilon\ll 1$, changing the commitment from $\Gamma_1^*$ to $\Gamma_1'$ in this case will not affect $S_2$'s optimal signalling strategy. However, the revised commitment $\Gamma_1'$ has a higher total probability of $\hat{\omega}_1'$ comparing to the signal $\hat{\omega}_1$ under $\Gamma_1^*$. Given the inequality $\Expect{\PRIOR}{U_{S_1}(a^*_{\Gamma_1^*,\Gamma_2^*}(\hat{\omega}_1,\Signal{2}))}>\Expect{\PRIOR}{U_{S_1}(a^*_{\Gamma_1^*,\Gamma_2^*}(\bar{\omega}_1,\Signal{2}))}$ and the knowledge that both the optimal commitments of $S_2$ and the receiver's best response stay unchanged, $S_1$ will enjoy a higher expected utility under the the commitment $\Gamma_1'$, which contradicts the assumption that $\Gamma_1^*$ is the optimal commitment of $S_1$. Hence, one of conditions in Claim \ref{clm:4} must hold in this case.

\textbf{Case 2: \Sender{2} can elicit the collaborative state $\hat{\STATE}$ under the signal $\hat{\omega}_1'$.}\\
This case occurs when \Sender{2}'s private information set $I_2\ni \hat{\STATE}$ contains no state that belongs to \Sender{1}'s information set $I_1$, i.e., $\StateX{i}\notin I_2, \forall \StateX{i}\in I_1, \hat{\STATE} \in I_2$. Therefore, even under the signal $\hat{\omega}_1'$ which mixes $\hat{\omega}_1$ with the signal containing the collaborative states, \Sender{2} can still persuade the receiver to take $\Act{\hat{\STATE}}$ with some probability (according to the prior). Besides, persuading the receiver to take $\Act{\hat{\STATE}}$ benefits both senders. However, after maximizing the signal \Sender{2} used to persuade the action $\Act{\hat{\STATE}}$, \Sender{2}'s optimal signalling strategy conditional on $\Act{\hat{\STATE}}$ not sent is the same as her optimal signalling strategy while receiving $\hat{\omega}_1$. Therefore, \Sender{2}'s optimal signalling changes by a small amount (because $\epsilon\ll 1$). Now, since $\hat{\STATE}$ is a collaborative state, \Sender{2}'s adjustment on her optimal signalling strategy from $\hat{\omega}_1$ under $\Gamma_1^*$ to $\hat{\omega}_1'$ under $\Gamma_1'$ not only benefits her but also benefits \Sender{1}. Moreover, after maximizing the probability of the signal persuading $\Act{\hat{\STATE}}$, the \Sender{2}'s optimal signalling strategy stays unchanged (reduced to the case 1 scenario). Hence, \Sender{1} enjoys a higher expected utility under the the commitment $\Gamma_1'$ in this case (because \Sender{2}'s optimal signalling strategy gives \Sender{1} a higher utility than the utility in case 1, and \Sender{1} enjoys a higher expected utility in case 1).

In both cases, \Sender{1}'s expected utility increases. This violates the assumption that \Sender{1} has the highest utility in $\hat{\omega}_1$ when condition 2 is violated. Hence, one of the conditions in Claim \ref{clm:4} must hold.

\end{proof}

\subsection{Proof of Theorem \ref{lem:suffcon}}
First, let us review the statements in condition 1 for the readability of the proof.

Condition 1:  There exist two parameters $\alpha>\beta>0$ and a signalling strategy $\hat{\Gamma}_2$ satisfying the following conditions:
    \begin{enumerate}
        \item[(a)] $G(p,\Info{1}{x}+\alpha \Info{1}{y})=G(p,\Info{1}{x}+\beta \Info{1}{y})$,
        \item[(b)] Let $\Signal{1}^\alpha,\Signal{1}^\beta$ be two mock signals of \Sender{1} such that $\Prob{\Info{1}{x}|\Signal{1}^\alpha}=\frac{1}{1+\alpha}$,
    $\Prob{\Info{1}{y}|\Signal{1}^\alpha}=\frac{\alpha}{1+\alpha}$,
    $\Prob{\Info{1}{x}|\Signal{1}^\beta}=\frac{1}{1+\beta}$, and 
    $\Prob{\Info{1}{y}|\Signal{1}^\beta}=\frac{\beta}{1+\beta}$, then
    \begin{align*}
        \hat{\Gamma}_2(p,\Signal{1}^\beta)\in G(p,\Info{1}{x}+\beta \Info{1}{y})~\text{and}~\hat{\Gamma}_2(p,\Signal{1}^\alpha)\notin G(p,\Info{1}{x}+\alpha \Info{1}{y}),
    \end{align*}
        \item[(c)] $\Expect{p}{U_{S_1}|\Signal{1}^\alpha,\hat{\Gamma}_2}<  \frac{\Prob{\Info{1}{x}}+\beta \Prob{\Info{1}{y}}}{\Prob{\Info{1}{x}}+\alpha \Prob{\Info{1}{y}}}\Expect{p}{U_{S_1}|\Signal{1}^\beta,\hat{\Gamma}_2}
        +\frac{(\alpha-\beta)\Prob{\Info{1}{y}}}{\Prob{\Info{1}{x}}+\alpha \Prob{\Info{1}{y}}}\Expect{p}{U_{S_1}|\Info{1}{y},\hat{\Gamma}_2}$,
        \item[(d)] $\Expect{p}{U_{S_2}|\Signal{1}^\beta,G(p,\Info{1}{x}+\beta \Info{1}{y})}< \Expect{p}{U_{S_2}|\Info{1}{y},G(p, \Info{1}{y})}$.
    \end{enumerate}

Given (a), if \Sender{1} commits first and the two mock signals $\Signal{1}^\alpha,\Signal{1}^\beta$ described in (b) are used in \Sender{1}'s signalling strategy, $G(p,\Info{1}{x}+\alpha \Info{1}{y})=G(p,\Info{1}{x}+\beta \Info{1}{y})$ states that the set of \Sender{2}'s optimal signalling strategies under $\Signal{1}^\alpha$ is the same as the set of \Sender{2}'s optimal signalling strategies under $\Signal{1}^\beta$. In other words, $\Signal{1}^\alpha$ and $\Signal{1}^\beta$ are two different mixing schemes using \Sender{1}'s information sets $\Info{1}{x}$ and $\Info{1}{y}$ that elicit the same response from \Sender{2}. According to the statement of condition 2, i.e., $\min_{g\in G(p,\Info{1}{x})}\Expect{p,g}{U_{S_1}|\Info{1}{x}}> \max_{g'\in G(p,\Info{1}{y})}\Expect{p,g'}{U_{S_1}| \Info{1}{y}}$, when \Sender{2}'s action and the receiver's action both stay unchanged, \Sender{1} prefers signal $\Signal{1}^\alpha$ over signal $\Signal{1}^\beta$. The logic is the following: First, suppose the receiver takes action $a\in \Info{1}{x}$, sending signal $\Signal{1}^\alpha$ mixes a higher portion of $\Info{1}{y}$ with $\Info{1}{x}$ than sending signal $\Signal{1}^\beta$. This implies that the receiver takes action $a\in \Info{1}{x}$ with a higher total probability. Based on the inequality $\min_{g\in G(p,\Info{1}{x})}\Expect{p,g}{U_{S_1}|\Info{1}{x}}> \max_{g'\in G(p,\Info{1}{y})}\Expect{p,g'}{U_{S_1}| \Info{1}{y}}$, increasing the total probability of a set of actions $a\in \Info{1}{x}$ by reducing the same amount of probability on a set of actions $a'\in \Info{1}{y}$ benefits \Sender{1}; second, suppose the receiver takes action $a'\in \Info{1}{y}$ under both signal $\Signal{1}^\alpha$ and signal $\Signal{1}^\beta$, \Sender{1} still prefers using the signal $\Signal{1}^\alpha$ when condition 2 holds, since under this circumstance signal $\Signal{1}^\alpha$ uses less probability mass of states in $\Info{1}{x}$ than signal $\Signal{1}^\beta$. In short, combining the statements in condition 1.(a) and condition 2, we know \Sender{1} prefers sending signal $\Signal{1}^\alpha$ over signal $\Signal{1}^\beta$. Furthermore, when we consider the statements in condition 1.(c) and 2, the inequality in 1.(c) implies \Sender{1} can persuade the receiver to take action $a\in \Info{1}{x}$ under signal $\Signal{1}^\beta$. (Otherwise, the inequality is violated since the receiver will take the same action $\Act{a}'\in \Info{1}{y}$ under signal $\Signal{1}^\alpha$, signal $\Signal{1}^\beta$, and the signal which reveals $\Info{1}{y}$.) Therefore, combining statements in condition 1.(a), condition 1.(c), and condition 2, we know \Sender{1} can persuade the receiver taking actions $a\in \Info{1}{x}$ under signal $\Signal{1}^\alpha$ and signal $\Signal{1}^\beta$, and \Sender{1} prefers sending signal $\Signal{1}^\alpha$.

Now let's explore \Sender{2}'s preference on signals/information sets. Given condition 1.(d), the inequality $\Expect{p}{U_{S_2}|\Signal{1}^\beta,G(p,\Info{1}{x}+\beta \Info{1}{y})}< \Expect{p}{U_{S_2}|\Info{1}{y},G(p, \Info{1}{y})}$ states that \Sender{2} prefers a (pure) signal eliciting \Sender{1}'s information $\Info{1}{y}$ over a mixed signal $\Signal{1}^\beta$. Moreover, condition 1.(a) states that the set of \Sender{2} optimal signalling strategies under $\Signal{1}^\beta$ is the same as the set of \Sender{2} optimal signalling strategies under $\Signal{1}^\alpha$. Thus, with $\alpha>\beta$ and the inferred fact that \Sender{1} can persuade the receiver taking $a\in \Info{1}{x}$ under signal $\Signal{1}^\alpha$ and signal $\Signal{1}^\beta$, we can derive the inequality $\Expect{p}{U_{S_2}|\Signal{1}^\alpha,G(p,\Info{1}{x}+\alpha \Info{1}{y})}< \Expect{p}{U_{S_2}|\Info{1}{y},G(p, \Info{1}{y})}$ from $\Expect{p}{U_{S_2}|\Signal{1}^\beta,G(p,\Info{1}{x}+\beta \Info{1}{y})}< \Expect{p}{U_{S_2}|\Info{1}{y},G(p, \Info{1}{y})}$. Combining the two inequalities derived above, we can write the following inequality: \begin{align}
    \max\big\{\Expect{p}{U_{S_2}|\Signal{1}^\alpha,G(p,\Info{1}{x}+\alpha \Info{1}{y})},\Expect{p}{U_{S_2}|\Signal{1}^\beta,G(p,\Info{1}{x}+\beta \Info{1}{y})}\big\}< \Expect{p}{U_{S_2}|\Info{1}{y},G(p, \Info{1}{y})}. \label{ineq:1}
\end{align}
Given the inequality (\ref{ineq:1}), we can conclude that \Sender{2} prefers a separation of \Sender{1}'s information sets \Info{1}{x} and \Info{1}{y}. 

Now, condition 1.(b) and condition 1.(c) guarantee the existence of a signalling strategy $\hat{\Gamma}_2$ satisfying the following three conditions:
\begin{itemize}
    \item $\hat{\Gamma}_2(p,\Signal{1}^\beta)\in G(p,\Info{1}{x}+\beta \Info{1}{y})$,
    \item $\hat{\Gamma}_2(p,\Signal{1}^\alpha)\notin G(p,\Info{1}{x}+\alpha \Info{1}{y})$,
    \item $\Expect{p}{U_{S_1}|\Signal{1}^\alpha,\hat{\Gamma}_2}<  \frac{\Prob{\Info{1}{x}}+\beta \Prob{\Info{1}{y}}}{\Prob{\Info{1}{x}}+\alpha \Prob{\Info{1}{y}}}\Expect{p}{U_{S_1}|\Signal{1}^\beta,\hat{\Gamma}_2}
        +\frac{(\alpha-\beta)\Prob{\Info{1}{y}}}{\Prob{\Info{1}{x}}+\alpha \Prob{\Info{1}{y}}}\Expect{p}{U_{S_1}|\Info{1}{y},\hat{\Gamma}_2}$.
\end{itemize}
Therefore, $\hat{\Gamma}_2$ is a signalling strategy that can threaten $\Sender{1}$. Given $\hat{\Gamma}_2$, \Sender{1} is better off by committing to signalling strategies using signal $\Signal{1}^\beta$ instead of signal $\Signal{1}^\alpha$, guaranteed by the condition 1.(c). This increases the total probability of the signal which elicits \Sender{1}'s information $\Info{1}{y}$, and then increases \Sender{2}'s expected utility (because \Sender{2} prefers separation over mixture on the information sets $\Info{1}{x}$ and $\Info{1}{y}$). In summary, given conditions (a)-(d) in condition 1 and the inequality in condition 2, \Sender{2} has a conflict of interest with \Sender{1} on the mixture of $\Info{1}{x}$ and $\Info{1}{y}$. However, \Sender{2} is willing to (partially) collaborate with \Sender{1} under $\Info{1}{x}$ when \Sender{2} has no alternatives (because $G(p,\Info{1}{x}+\alpha \Info{1}{y})=G(p,\Info{1}{x}+\beta \Info{1}{y})$).

At a high level, a collaboration under $\Info{1}{x}$ benefits both \Sender{1} and \Sender{2} no matter who commits first. However, \Sender{1} and \Sender{2} hold different opinions on the scale of collaboration, where \Sender{1} want to maximize the total probability of collaboration, but \Sender{2} only wants to collaborate when she has no alternative (signalling strategies that are incentive-compatible to the receiver). Given the first two conditions, if the prior distribution supports it, then both senders will collaborate on $\Info{1}{x}$. Condition 1.(b) and 1.(c) guarantees that \Sender{2} can construct a credible threat to reduce the total probability of collaboration when she commits first, and this makes the commitment order matter. Finally, condition 3 serves as a regularity condition which ensures that the prior distribution supports the collaboration and the (potential) credible threat. Hence, when three conditions in Theorem \ref{lem:suffcon} hold, the commitment order matters.

\subsection{Proof of Theorem \ref{lem:necessary}}
The idea of this proof is to prove that the commitment order may matter if one of the conditions is not violated.

First, if the receiver's tie-breaking rule is belief-dependent, the sender who commits later can tailor the posterior beliefs to make the tie-breaking rule favor her. Hence, unless all tie-breaking decisions made by the receiver are indifferent to both senders \Sender{1} and \Sender{2}, both prefer to commit last. In order to verify whether the commitment order matters when condition 1 holds, we assume that condition 2 is violated.

Suppose the receiver's tie-breaking rule is belief independent; we then prove that satisfying condition 1 may make the commitment order matters. The idea of proving the above statement is to construct the minimum requirements such that (partial) collaboration and credible threats can both occur. We will illustrate the conditions required for a (partial) collaboration and a credible threat below, respectively.

First, we derive the minimum requirement of a (partial) collaboration. \Sender{1} and \Sender{2} will collaborate when their preferences (ordering) on at least a pair of states align, called $\StateX{\alpha}$ and $\StateX{\beta}$ hereafter for simplicity of representation. Moreover, if collaboration occurs in $\StateX{\alpha}$ and $\StateX{\beta}$, \Sender{1} must be unable to distinguish $\StateX{\alpha}$ and $\StateX{\beta}$ using her private information, i.e.,$\exists! \Info{1}{x}\ni \theta_{\alpha}, \theta_{\beta}$. Otherwise, \Sender{1} will separate these two states herself (if it is best for her) and no collaboration occurs. Given assumption 4 in this chapter, \Sender{2} can learn the true state via her private information when \Sender{1} reveals her information truthfully. This demands $\StateX{\alpha}$ and $\StateX{\beta}$ must belong to different information sets of \Sender{2}. To make a collaboration on $\StateX{\alpha}$ and $\StateX{\beta}$ possible, \Sender{1} and \Sender{2} must have aligned preference orders on these two states. Without loss of generality, we assume $U_{S_1}(\Act{\alpha})>U_{S_1}(\Act{\beta})$ and $U_{S_2}(\Act{\alpha})>U_{S_2}(\Act{\beta})$. Collecting the points together, to construct a (partial) collaboration, we need the following conditions to hold:
\begin{itemize}
    \item $\exists! \Info{1}{x}\ni \theta_{\alpha}, \theta_{\beta}$
    \item $\exists! \Info{2}{x}\ni \theta_{\alpha}$ and $\exists!\Info{2}{y}\ni \theta_{\beta}$ such that $\Info{2}{x}\neq \Info{2}{y}$.
    \item $U_{S_1}(\Act{\alpha})>U_{S_1}(\Act{\beta})$
    \item $U_{S_2}(\Act{\alpha})>U_{S_2}(\Act{\beta})$
\end{itemize}

Next, we derive the minimum requirements for a (potential) conflict of interest. Before the analysis, we pause to note that a conflict of interest is not guaranteed under our requirements, because we do not specify the utilities of \Sender{1} and \Sender{2} on the receiver's best-response action corresponding to other states within the same information set of \Sender{1} (if any). Precisely, when $\Info{1}{y}$ is the information set involved in a (potential) conflict of interest between senders, we do not specify $\Util{\Sender{1}}{\StateX{\tau}}{\Act{\tau}}$ and $\Util{\Sender{2}}{\StateX{\tau}}{\Act{\tau}}$ for every state $\StateX{\tau}\in \Info{1}{y}$ not belonging to the set of states where we construct a (potential) conflict of interest.

A (potential) conflict of interest arises when \Sender{1} and \Sender{2} prefer different mixed strategies on a set of \Sender{1}'s information sets. Since we are interested in the minimum requirements, we search for conflict of interests on a pair of information sets, called $\Info{1}{x}$ and $\Info{1}{y}$, where the information set $\Info{1}{x}$ is the same set where collaboration may occur. (Because we demand the occurrence of collaboration and a conflict of interests. If only collaborations occur, \Sender{1} and \Sender{2} can reach a consensus on their (collectively) optimal commitments. If only a conflict of interests occurs, \Sender{1} will commit to a maximin strategy, and \Sender{2} will commit to a minimax strategy. This is because \Sender{1} always sends signals before \Sender{2}.) 

Before constructing a (potential) conflict of interest, we assume $\StateX{\gamma}$ is a state that belongs to \Sender{1}'s information $\Info{1}{y}$. As presented earlier, a (potential) conflict of interest requires that \Sender{1} and \Sender{2} have difference preference on a mixture of $\Info{1}{x}$ and $\Info{1}{y}$. Here we assume\footnote{For the opposite conflict of interest, i.e., \Sender{1} prefers a separation over mixture and Sender{2} prefers a mixture over separation, the whole analysis is exactly analogous with a swap of senders' utility inequalities. The inequalities restricting \Sender{1} will now restrict \Sender{2}, and vice versa. In short, we just swap $\Sender{1}$ with $\Sender{2}$ in condition 1.(b).} \Sender{1} prefers a mixture over separation and \Sender{2} prefers a separation over mixture.

Since \Sender{1} prefers a mixture of $\Info{1}{x}$ and $\Info{1}{y}$, there must exists a state $\Sender{\gamma} \in \Info{1}{y}$ such that $U_{S_1}(\Act{\alpha})>U_{S_1}(\Act{\gamma})$, where $\StateX{\alpha} \in \Info{1}{x}$ as used earlier. Similarly, when \Sender{2} prefers a separation over a mixture, there must exists a state $\Sender{\gamma} \in \Info{1}{y}$ such that $U_{S_2}(\Act{\alpha})<U_{S_2}(\Act{\gamma})$. To avoid a violation of assumption 1, i.e., each sender only has partial information about the state of the world, in a ternary state space corner case, we let $\StateX{\gamma}$ belongs to either $\Info{2}{x}$ or $\Info{2}{y}$. To sum up, to construct a minimum (potential) conflict of interest op top of a partial collaboration, the following additional conditions hold:
\begin{itemize}
    \item $U_{S_1}(\Act{\alpha})>U_{S_1}(\Act{\gamma})$
    \item $U_{S_2}(\Act{\alpha})<U_{S_2}(\Act{\gamma})$
    \item Either $\StateX{\gamma} \in \Info{2}{x}$ or $\StateX{\gamma} \in \Info{2}{y}$.
\end{itemize}
Now, we have constructed the minimum requirement of collaboration and the minimum requirement of a conflict of interest. The collection of information structure requirements is the statement of condition 1.(a) in Theorem \ref{lem:necessary}, and the requirements on \Sender{1}'s and \Sender{2}'s utilities are summarized in the inequalities in condition 1.(b). Therefore, when condition 1 is satisfied, the commitment order may matter because of the co-occurrence of collaboration and a conflict of interest. Thus, when both conditions are violated, the commitment order does not matter.

\subsection{Utility calculation of Example 3.1, Sender 1 commits first}\label{app:utilcal3.1.1}
\begin{eqnarray}
 &\Expect{\PRIOR}{\UTIL{\Sender{1}}} &=(0.2+0.3\times \tfrac{2}{3}+0.4\times \tfrac{2}{3}\times \tfrac{3}{4} )\times 2+0=1.2 \\
 & \Expect{\PRIOR}{\UTIL{\Sender{2}}} &=(0.2+0.3\times \tfrac{2}{3}+0.4\times \tfrac{2}{3}\times \tfrac{3}{4} )\times 2+\tfrac{1}{3}\times (0.3+0.4\times\tfrac{3}{4})\times 3 \nonumber\\ &&=1.2+0.6=1.8 \\
 & \Expect{\PRIOR}{\UTIL{R}} &=0.2+0.3\times \tfrac{1}{3}+0.4 \times \tfrac{1}{3} \times \tfrac{1}{4}=\tfrac{1}{3}
\end{eqnarray}
\subsection{Utility calculation of Example 3.1, Sender 2 commits first} \label{app:utilcal3.1.2}
\begin{eqnarray}
 &\Expect{\PRIOR}{\UTIL{\Sender{1}}}&=(0.2+0.3\times \tfrac{1}{2}+0.4\times \tfrac{1}{2})\times 2+0=1.1 \\
 &\Expect{\PRIOR}{\UTIL{\Sender{2}}}&=(0.2+0.3\times \tfrac{1}{2}+0.4\times \tfrac{1}{2})\times 2+\frac{1}{2}\times (0.3+0.4\times\tfrac{3}{4})\times 3 \nonumber\\ &&=1.1+0.9=2 \\
  &\Expect{\PRIOR}{\UTIL{R}}&= 0.2+0.3\times \tfrac{1}{2}+0.4\times\tfrac{1}{2}\times\tfrac{1}{4}=0.4 
\end{eqnarray}    
\subsection{Utility calculation of Example 4.1, Sender 1 commits first} \label{app:utilcal4.1.1}
\begin{eqnarray}
 &\Expect{\PRIOR}{\UTIL{\Sender{1}}}&=(0.05+0.05+0.03+0.05+0.005+0.05+0.03)\times 5 \nonumber\\&& + (0.09\times 0.54+0.15\times 0.9+0.25\times 0.54+0.15\times 0.9)\times 2\nonumber\\&&
 +(0.25\times 0.4+0.09\times 0.46)+0.03\times 2 =2.4336 \\
 &\Expect{\PRIOR}{\UTIL{\Sender{2}}}&=(0.05+0.05+0.03+0.05+0.005+0.05+0.03)\times 5 \nonumber\\&& + (0.09\times 0.54+0.15\times 0.9+0.25\times 0.54+0.15\times 0.9)\times 2\nonumber\\&&
 +(0.015\times 4+0.09\times 0.06)\times 3+0.03\times 2\times 4+0.02\nonumber\\&& =2.6884 \\
  &\Expect{\PRIOR}{\UTIL{R}}&= 0.05+0.02\times 0.75+0.2\times 0.75\times 0.9+0.12\times 0.75\times 0.36\nonumber\\&&+0.35\times\tfrac{1}{7}\times 0.4+0.21\times\tfrac{1}{7}\approx0.286.
\end{eqnarray}    
\subsection{Utility calculation of Example 4.1, Sender 2 commits first} \label{app:utilcal4.1.2}
\begin{eqnarray}
 &\Expect{\PRIOR}{\UTIL{\Sender{1}}}&=(0.05+0.05+0.03+0.05+0.005+0.05+0.03)\times 5 \nonumber\\&& + (0.09\times 0.9+0.15\times 0.9+0.15\times 0.9+0.09\times 0.9)\times 2\nonumber\\&&
 +(0.15\times 0.6+0.09) =2.207 \\
 &\Expect{\PRIOR}{\UTIL{\Sender{2}}}&=(0.05+0.05+0.03+0.05+0.005+0.05+0.03)\times 5 \nonumber\\&& + (0.09\times 0.9+0.15\times 0.9+0.15\times 0.9+0.09\times 0.9)\times 2\nonumber\\&&
 +(0.015\times 3+0.009\times 2)\times 3+0.09\times 2\times 4+0.06 =3.158 \\
  &\Expect{\PRIOR}{\UTIL{R}}&= 0.05+0.02\times 0.75+0.2\times 0.75\times 0.9\nonumber\\&&+0.35\times\tfrac{3}{7}\times 0.4+0.21\times\tfrac{3}{7}=0.35.
\end{eqnarray}

\end{document}